\documentclass[final,5p]{elsarticle}

\usepackage[utf8]{inputenc}

\usepackage{graphicx}
\usepackage{subfig}
\usepackage{epstopdf}

\usepackage{amssymb}
\usepackage{amsmath}

\usepackage{multirow}
\usepackage{booktabs}

\usepackage{algorithm}
\usepackage{algpseudocode}

\usepackage{color}

\journal{Computer Physics Communications}

\begin{document}

\begin{frontmatter}

\title{A new {GPU} implementation for lattice-{Boltzmann} simulations on sparse geometries}

\author[a]{Tadeusz Tomczak\corref{author}}
\author[b]{Roman G. Szafran}

\cortext[author] {Corresponding author.\\\textit{E-mail address:} tadeusz.tomczak@pwr.wroc.pl}
\address[a]{Wrocław University of Science and Technology, Faculty of Electronics, Janiszewskiego 11/17, 50-370 Wrocław, Poland, e-mail: \emph{tadeusz.tomczak@pwr.wroc.pl}}
\address[b]{Wrocław University of Science and Technology, Faculty of Chemistry, C.K. Norwida 4/6, 50-373 Wrocław, Poland, e-mail: \emph{roman.szafran@pwr.wroc.pl}}

\begin{abstract}
We describe a high-performance implementation of the lattice Boltzmann method (LBM) for sparse 3D geometries on graphic processors (GPU).
The main contribution of this work is a data layout that allows to minimise the number of redundant memory transactions during the propagation step of LBM.
We show that by using a uniform mesh of small three-dimensional tiles and a careful data placement it is possible to  utilise more than 70\% of maximum theoretical GPU memory bandwidth for D3Q19 lattice and double precision numbers.
The performance of our implementation is thoroughly examined and compared with other GPU implementations of LBM.
The proposed method performs the best for sparse geometries with good spatial locality.
\end{abstract}


\end{frontmatter}

\section{Introduction}
\label{}

The lattice Boltzmann method (LBM) is a versatile and highly parallel approach where the discrete Boltzmann transport equation is solved in the velocity or moment $\mathbf{R}^n$ space to obtain the time-dependent fluid velocity distributions. 
Due to a straightforward LBM parallelization,
many researchers have explored the area of its implementations on graphics processors (graphic processing units, GPU).
A thorough review is presented in \cite{Mawson:Mem2014}.
A few typical optimisation techniques may be mentioned: combining all computations into the single kernel to minimise memory bandwidth usage, use of structure of arrays (SoA) instead of array of structures (AoS) data types to avoid uncoalesced memory transactions, an adaptation of all computations to the single precision floating point format to minimise a register pressure and maximise occupancy \cite{Valero-Lara:Acc2014,Valero-Lara:Acc2015}, use of two copies of data to avoid race conditions, methods decreasing memory usage \cite{Valero-Lara:Het2017, Valero-Lara:Lev2016, Valero-Lara:ANo2015, Latt:How2007}, and many others.

However, the above optimisations concern dense geometries.
The number of GPU implementations for sparse geometries (with many solid nodes) is much lower and they offer significantly worse performance.
This can hinder the usage of GPU LBM implementations in many areas where dense geometries are not applicable.
Typical examples may be biomedical or porous media simulations 
\cite{Huang:Mul2014}.

Sparse geometries handling requires some form of storing information about the placement of non-solid nodes.
This information can not only increase memory usage, but also generate irregular memory access patterns that can significantly decrease performance.

In general, two indirect addressing solutions are used in GPU implementations of LBM for sparse geometries: the Connectivity Matrix (CM), shown in \cite{Bernaschi:AFl2010} with highly optimised version in \cite{Huang:Mul2014}, and the Fluid Index Array (FIA), presented in \cite{Nita:GPU2013}.
The connectivity matrix contains pointers to the neighbour nodes for each lattice direction.
The fluid index array is a kind of "bitmap" that contains $-1$ for every solid node in geometry or the pointer to non-solid node data.
Both solutions bring an overhead in memory usage, additionally the FIA implementation results in a low utilisation of memory bandwidth due to the irregular data access pattern.

LBM implementations for dense geometries often use data structures that allow placing values for neighbour nodes in consecutive memory locations. 
This results in the linear memory traversing during collision and propagation and a lack of uncoalesced memory transfers, giving the high performance as a consequence.
For sparse implementations with indirect addressing, there is no guarantee that neighbour node values will be placed in neighbour memory locations.
This may result in an additional memory bandwidth usage caused by uncoalesced memory transactions containing values for unused nodes.
These uncoalesced memory transactions, together with transfers of additional pointers to non-solid nodes, results in a significant performance penalty for implementations based on indirect addressing.

The solution proposed in this work allows coping with this disadvantage of implementations for sparse geometries.
We present the GPU implementation of LBM where
the information about geometry sparsity is stored using the uniform grid of three-dimensional cubic tiles.
By partitioning the domain into small tiles the three main goals are achieved. 
First, we almost completely removed additional data that in indirect addressing are used as pointers to non-solid nodes.
Second, due to the fixed tile size, the values for nodes from the tile can be placed in memory in a way that guarantees a minimum amount of additional memory traffic and, as a consequence, the high performance of single tile processing.
 At the same time, a low memory usage and computational complexity for sparse geometries can be achieved by simply skipping tiles filled with only solid nodes. 
 Thus, the proposed solution allows simulating flows through irregular, sparse geometries with the performance close to efficient LBM implementation for dense geometries.

For the proposed data layout, we show both a detailed theoretical analysis and results of performance measurements for dense and sparse geometries.
Additionally, we analyse how the fixed tile size affects the overall performance.
Provided that the tiles contain a low number of solid nodes, the performance of our implementation remains high regardless of the geometry sparsity.

The structure of the paper is as follows.
In Section \ref{sec_background}, we briefly introduce the LBM method and define basic concepts in GPU programming.
Section \ref{sec_impl} contains the description of our implementation with the detailed analysis of introduced overheads.
The performance comparison with existing implementations and a verification procedure are presented in Section \ref{sec_results}.
Section \ref{sec_conclusions} contains conclusions.

\section{Background}
\label{sec_background}

\subsection{Graphic Processing Unit}

Today's GPUs are in fact mass-produced, easily programmable versatile variant of vector coprocessors.
In this work, we use a GPU compatible with the CUDA technology \cite{cuda7.5:2015}.
Selected parameters of CUDA GPU architectures are shown in Table \ref{tab_GPU}.
It can be seen that the successive generations offer not only a higher memory bandwidth but also their computational power for double precision increases faster than the memory bandwidth.

\begin{table}[htbp]
\caption{
	Approximate parameters of CUDA capable GPUs.
	We consider only models with maximum double precision performance dedicated to the high performance computing.
	The last row contains parameters of a high-end general purpose CPU.
}
\label{tab_GPU}
\centering
\begin{tabular}{c c c c}
\hline
Architecture & Launch & Bandwidth    & DP FLOP/byte   \\
             &  year  &  [GB/s]      &             \\
\hline 
Tesla        &  2008  &  $\sim 100$  &  $\sim 0.8$ \\
Fermi        &  2009  &  $\sim 150$  &  $\sim 3.5$ \\
Kepler       &  2012  &  $\sim 250$  &  $\sim 5.5$ \\
Pascal       &  2016  &  $\sim 720$  &  $\sim 7.4$ \\
\hline
Broadwell    &  2016  &  $\sim 80$   &  $\sim 8$   \\
\hline
\end{tabular}
\end{table}

Since this paper investigates memory bandwidth optimisations, thus we briefly recall characteristics of the memory system in CUDA GPUs.
Detailed information can be found in \cite{cuda7.5:2015} and in the specific architecture tunning guides distributed with the CUDA platform.

GPU memory hierarchy contains the off-chip DRAM \emph{global} memory (limited to 1--32 GB depending on GPU), up to two levels of on-chip cache memory (there may be separate caches for read-only access), shared memory and registers.
A single transaction with the global memory requires a few hundreds of GPU clock cycles, thus memory transactions for threads from the same warp can be \emph{coalesced} into one if defined restrictions are met.
Additionally, limited software control over a cache usage is available.

Coalescing rules depend on GPU architecture generation.
For Fermi class, the single global memory transaction can contain 128 or 32 bytes depending on whether L1 cache memory with 128 byte-wide lines is used.
For the first generation Kepler devices (up to GK110 processor), the size of global memory transaction is fixed at 32 bytes (L1 cache is used only for local memory).
The second generation Kepler processors (GK110B and GK2xx devices) can optionally use 128-byte global memory transactions.
For Maxwell architecture, the transaction size also depends on version.
Current Pascal devices use 32-byte global memory transactions regardless whether L1 cache is used or not.
Thus, in further considerations, we assume that the single memory transaction contains 32 bytes because only this size is supported by our machine -  GTX Titan card with Kepler architecture GK110 processor.
Moreover, for irregular memory access patterns, the smaller transactions result in higher bandwidth utilisation due to lower overhead as presented in \cite{cuda7.5:2015}.

\subsection{Lattice-Boltzmann method}

\begin{figure}[htbp]
\centering
\includegraphics[width=\columnwidth]{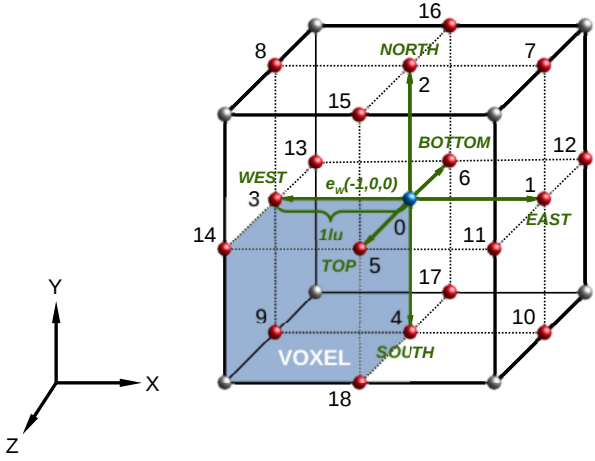}
\caption{
	A grid composed of 8 unit volumes (voxels). 
	D3Q19 lattice nodes: blue - central node 0, red - streaming nodes (1-18), gray - not directly linked with central node. 
	The convention of lattice directions naming: west (W) direction vector $\boldsymbol{e}_W$ or $\boldsymbol{e}_3 = (-1,0,0)$ of length 1 lu (lattice unit) points out from the central node in direction 3.
}
\label{fig_directions}
\end{figure}

In the LBM, as in the conventional CFD method, the geometry, initial and boundary conditions must be specified to solve the initial value problem.
The values of macroscopic quantities can not be directly imposed on boundaries and the initial and boundary values of distribution functions have to be calculated from the known values of the macroscopic quantities after conversion to the lattice unit system using dimensionless numbers (e.g. Reynolds number) \cite{Kruger:The2016}.
For the inlet of the domain we used Zou-He \emph{velocity} boundary condition and \emph{constant pressure} condition for the outlet \cite{Zou:OnP1997}.
At the walls we applied mixed boundary conditions: \emph{velocity 0} and \emph{bouce-back} \cite{Chen:OnB1996}. 
The computational domain is uniformly partitioned and computational nodes are placed in corners of adjacent cells (voxels), giving the lattice.
In this study, the D3Q19 lattice structure is used, where D$_b$ represents the space dimension and Q$_n$ the lattice links number.
Detailed description of the LB method one can find in recently published books \cite{Kruger:The2016}, \cite{Guo:Lat2013} or \cite{Mohamad:Lat2011}. 
In this work, we use the lattice directions naming scheme shown in  Fig. \ref{fig_directions}.

The transport equation of the lattice-Boltzmann meth\-od (LBE) can be written for each direction in the momentum space as follows:
\begin{equation}
	\frac{ \partial f_i}{ \partial t} + \boldsymbol{c}_i \cdot \nabla f_i = \Omega_i
	,
	\label{equ_lbm_1}
\end{equation}
where $f_i$ represents the probability distribution function (PDF) along the $i$ lattice direction,
$\delta_x$ and $\delta_t$ are the lattice spacing and the lattice time step (usually assumed to be equal 1),
$\frac{\delta_x}{\delta_t}$ is the lattice speed (also usually assumed to be equal 1 lu),
$\boldsymbol{e}_i$ is the unit vector along the $i$ lattice direction,
and $\boldsymbol{c}_i = \frac{\delta_x}{\delta_t} \boldsymbol{e}_i$ is the lattice velocity vector in the velocity space along the $i$ lattice direction,
and $\Omega_i$ is the collision operator.

The collision operator can be approximated with the most popular LBGK Bhatnagar-Gross-Krook model \cite{Bhatnagar:AMo1954} 
\begin{equation}
	\Omega_{LBGK} = - \frac{1}{\tau} \left( f_i - f_i^{eq} \right)
	,
\end{equation}
where $\tau$ is the relaxation time in LB units related to the lattice fluid viscosity 
and $f^{eq}_i$ is the momentum equilibrium distribution function (EDF) along the $i$ lattice direction given by the following formula for the quasi-compressible model of fluid flow \cite{Qian:Lat1992,Succi:The2001}:
\begin{equation}
	f^{eq}_i = \omega_i \rho \left( 1 + \frac{ \boldsymbol{c}_i \cdot \boldsymbol{u} }{ c_s^2 } + \frac{ \left( \boldsymbol{c}_i \cdot \boldsymbol{u} \right)^2 }{ 2  c_s^4 } - \frac{\boldsymbol{u}^2}{2  c_s^2}  \right)
	\label{equ_feq_qcompr}
\end{equation}
and for the incompressible model \cite{Zou:Aim1995}:
\begin{equation}
	f^{eq}_i = \omega_i  \left( \rho + \frac{ \boldsymbol{c}_i \cdot \boldsymbol{u} }{ c_s^2 } + \frac{ \left( \boldsymbol{c}_i \cdot \boldsymbol{u} \right)^2 }{ 2 c_s^4 } - \frac{\boldsymbol{u}^2}{2  c_s^2}  \right)
	\label{equ_feq_incompr}
\end{equation}
where $c_s$ is the lattice speed of sound that is a lattice constant: $c_s = \frac{1}{\sqrt{3}}$; 
$\boldsymbol{u}$ is the macroscopic fluid velocity vector expressed in LB units; 
$\omega_i$ is a weighting scalar for the $i$ lattice direction;
$\rho$ is a fluid density expressed in LB units: $\rho = \sum\nolimits_i f_i$ that is related to pressure expressed in LB units: $p = \frac{\rho}{3}$.
The weighting factors of D3Q19 lattice are: 
$\omega_i = 1/3$ for $\boldsymbol{e}_0$ direction vector;
$\omega_i = 1/18$ for $\boldsymbol{e}_1 \ldots \boldsymbol{e}_6$ direction vectors;
$\omega_i = 1/36$ for $\boldsymbol{e}_7 \ldots \boldsymbol{e}_{18}$ direction vectors.
It can be seen that both EDF (Eqn. (\ref{equ_feq_qcompr}) and (\ref{equ_feq_incompr})) are very similar except that the density is decoupled from the momentum in Eqn. (\ref{equ_feq_incompr}).

The macroscopic velocity for the quasi-compressible model can be determined from \cite{Succi:The2001}
\begin{equation}
	\boldsymbol{u} = \frac{1}{\rho} \sum\limits_i \boldsymbol{c}_i  f_i
\end{equation}
and for the incompressible model from \cite{He:Lat1997}
\begin{equation}
	\boldsymbol{u} = \sum\limits_i \boldsymbol{c}_i  f_i
	.
\end{equation}
Integrating Eqn. (\ref{equ_lbm_1}) from $t$ to $t + \delta_t$ along the $i$ lattice direction and assuming that the collision term is constant during the interval, we can obtain 
discretized in time form of LBGK equation
\begin{equation}
	\underbrace{f_i \left( \boldsymbol{r} + c_i \delta_t, t + \delta_t \right) - f_i(\boldsymbol{r}, t)}_{Streaming} = 
	\underbrace{\frac{ \delta_t}{\tau} \left[ f^{eq}_i (\boldsymbol{r},t) - f_i(\boldsymbol{r},t) \right] } _{Collision}
	,
	\label{equ_lbm_bgk}
\end{equation}
where $\boldsymbol{r}$ is a position vector in the velocity space.

The therm on the left hand side is known as the streaming step, the latter represents the collision step. 
These two steps are repeated sequentially during the simulation giving velocity, density and pressure distributions at each time step.

The LBGK uses the single relaxation time to characterise the collision effects. 
However, physically, these rates should be different during collision processes. 
To overcome this limitation, a collision matrix with different eigenvalues or multiple relaxation times can be used. 
The LBM with an LBMRT collision operator can be expressed as \cite{Lallemand:The2000}:
\begin{equation}
	\underbrace{f_i \left( \boldsymbol{r} + c_i \delta_t, t + \delta_t \right) - f_i(\boldsymbol{r}, t)}_{Streaming} = 
	\underbrace{ \boldsymbol{A} \left[ \boldsymbol{f}^{eq} - \boldsymbol{f} \right] } _{Collision}
	,
	\label{equ_lbm_mrt}
\end{equation}
where $\boldsymbol{A}$ is the collision matrix.
The LBMRT model has been attracting more attention recently due to the higher stability than that of the LBGK model. 
In our study, we use both above mentioned collision models in incompressible and quasi-compressible variants to compare their efficiency.

The hydrodynamic equations derived from the LBE (Eqn. (\ref{equ_lbm_1})) are the compressible Navier-Stokes equations, but the intrinsic low-Mach-number limit of classical LBE indicates that it is limited to weakly compressible flows only.
Therefore, the LBE can be viewed as special artificial compressibility method for incompressible flows, the so-called quasi-compressible model of fluid flow \cite{He:Com2002}.
The compressible effects may lead to significant errors in case of incompressible fluid flow simulations (for example liquids) or for high Mach number flows of gases.
In our simulations, to minimize the compressibility error when the quasi-compressible model is used to simulate liquid flows, values of the relaxation time $\tau$ and the grid density were chosen to fulfill condition: $u_{max} \leq 0.05$ \cite{Kruger:The2016}, where $u_{max}$ is a maximal macroscopic velocity noticed in the lattice.
In order to reduce compressible effects in LBE, some modifications of the standard models have been developed for low-Mach-number flows.
Zou and coworkers \cite{Zou:Aim1995} proposed the incompressible version of standard LBGK model Eqn. (\ref{equ_feq_incompr}).
In this model the compressible effects are removed totally at steady states, however, for unsteady flows, some errors still exist \cite{Guo:Lat2013}.

\subsection{LBM performance limits}
\label{sect_performance_model}
\label{sect_performance_limits}

To compare different implementations, we employ a widely used, simple computational complexity model based on \cite{Wellein:OnT2006} that allows for computing the minimum memory usage, the memory bandwidth and the computational performance.
The computed complexity is used to show that GPU implementations of LBM are often bandwidth-bound on current hardware.
We also use the model to calculate the idealised GPU memory bandwidth utilisation equal the ratio of real performance to the light-speed performance defined in \cite{Hager:Int2010} for bandwidth-bound implementations.
The computed bandwidth utilisation allows to estimate how close is our implementation to the "ideal" and to detect redundant operations that reduce performance.
Also, according to \cite{Hager:Int2010}, it may be treated as a factor used to compare different LBM implementations (provided that they are bandwidth-bound).

Since we are interested in peak performance, in the considerations that follow below we ignore boundary nodes.
Our implementation is based on the pull-approach presented in \cite{Valero-Lara:Acc2015, Rinaldi:ALa2012},
thus we also assume that the computational unit (CPU or GPU) does computations for a single node in a single LBM iteration (streaming and collision) in three stages: data read from the main memory, computations, data write to the main memory (the data for a single lattice node is buffered in registers and in the internal memory, e.g. shared/cache memory).
In our implementation the kernel is responsible for the full update of all geometry data (all $f_i$ values) for a single LBM time step iteration (collision, propagation and boundary computations are \emph{fused} into a single kernel).
Thus, the simplest version of simulation can contain only subsequent launches of the same kernel, of course after transferring geometry data to GPU memory at the beginning.

According to \cite{Wellein:OnT2006}, we assume that for each lattice node the only values stored in the main memory are $f_i$ functions.
Since the previous work has shown that usually LBM implementations are badnwidth-bound, the other values (e.g. velocity or density) can be computed internally, and dropped after use without transferring them from/to the main memory.

Let $q$ denote the number of functions $f_i$ and $n_d$ denote the number of bytes for storing a single $f_i$ value (e.g. 4 bytes for a single precision floating point, 8 bytes for double precision).
Thus, the minimum number of memory bytes needed for single node datum is
\begin{equation}
	M_{node} = q \cdot n_d ~~\mathit{[bytes]}
	\label{equ_mnode}
	.
\end{equation}

Many LBM implementations also use some additional data, but in this model, these data are treated as an overhead.
Fortunately, this overhead is almost negligible - usually, the largest part contains an additional field per node used to store information about the type of node (solid, fluid, kind of boundary condition).
This information can be efficiently encoded on single bytes (even a few bits for simple cases), which occupy less than 1\% of $M_{node}$ defined by Eqn. (\ref{equ_mnode}) for D3Q19 lattice arrangement.
Moreover, the node type field can be skipped if nodes are partitioned into sets containing nodes of the same type.

For a single LBM iteration, the new $f_i$ values are computed based on the node type and the previous $f_i$ values.
The newly computed $f_i$ values are stored in the memory.
Since each $f_i$ value is transferred twice (once read and once stored) then the minimum number of bytes transferred per one LBM iteration for a single node is
\begin{equation}
	B_{node} = 2 \cdot q \cdot n_d ~~ \mathit{[bytes]}
	\label{equ_bnode}
	.
\end{equation}

In theory, the value from Eqn. (\ref{equ_bnode}) can be further decreased.
It was shown in \cite{Pohl:Opt2003} that it is possible to do even a few time steps per one data read/write from the main memory by using the \emph{loop blocking} technique.
The general idea is to divide the domain into blocks that can be copied to the processor internal memory (cache and registers) and to do as much as possible computations for the cached nodes.
After updating some fragment of the cached block to the time step $t+1$, the computations for the next time step ($t+2$) can be started for all nodes that have neighbours valid for time $t+1$.
By carefully arranging the order in which nodes are processed, few time steps can be calculated without communication with the main memory.
However, according to \cite{Wellein:OnT2006} this method was difficult to apply to more complex boundary condition models.

\begin{table*}[htbp]
	\caption
	{
		Computational complexity of LBM operations for single fluid node, D3Q19 lattice and double precision values.
		First four rows show complexity for complete collision and $v,\rho$ computations.
		The last three rows shows complexity of separate $v,\rho$ computations - for quasi-compressible fluid model the computation of $v,\rho$ requires 3 additional divisions.
		Separate collision complexity can be calculated by subtracting complexity of $v,\rho$ computations from values from the first four rows.
		FMA (fused multiply-add) counts as two floating point operations.
		FSETP and FREC denote GPU instructions for floating point condition testing and reciprocal computing.
		FLOP/byte ratio is calculated assuming 304 bytes transferred per node (see Eqn. \ref{equ_bnode}).
	}
	\label{tab_FLOP}
	\centering
	\begin{tabular}{c c c c c c c c c c}
		\hline
		Operation                  & FADD & FMUL & FMA & FSETP & FREC  & \# instr.& FLOP  & FLOP/byte \\
		\hline
		LBGK incompressible         & 65	  & 21	& 109	  &  --	 & --	      & 195	 &   304 & 1,00  \\
		LBGK quasi-compressible      & 65	  & 39	& 166	  &  21	 & 6	      & 297	 &   463 & 1,52  \\
		LBMRT incompressible         & 324	& 40	& 329	  &  --	 & --	      & 693	 &  1022 & 3,36  \\
		LBMRT quasi-compressible      & 323	& 43	& 386	  &  21	 & 6	      & 780	 &  1165 & 3,83  \\
		\hline                                                                      
		$v,\rho$ incompressible    & 49	  & --  & --    & --   & --	      &  49	 &   49  &       \\
		$v,\rho$ quasi-compressible & 49	  & 15	& 57	  & 21	 & 6	      & 148	 &  205  &       \\
		FPU division               & --	  & 5	  & 19	  & 7	   & 2	      &  33	 &   52  &       \\
		\hline
	\end{tabular}
\end{table*}

The number of floating point operations (FLOP, not to be confused with floating point operations per second, FLOPS) is much more complex to determine.
A simple count of the operations resulting from a naive implementation of equations 
(\ref{equ_lbm_bgk}) and (\ref{equ_lbm_mrt})
gives values significantly larger than in real implementations.
Actually, many operations described by equations 
(\ref{equ_lbm_bgk}) and (\ref{equ_lbm_mrt})
can be skipped (e.g. multiplications by $-1, 0$ or $1$) or used a few times (e.g. partial sums of $f_i$).
Some of these optimisations can be detected very early (i.e. multiplications by $e_i$ constants equal to $-1, 0, 1$), but others are unpredictable since they are applied during the compilation and their use depends on many factors (optimisation level, registers usage constraints during compilation, machine capabilities - a number of registers etc.).
Thus, for numbers of computational operations we show only specific values obtained by a disassembling of a GPU binary code using \emph{nvdisasm} utility (we count only arithmetic operations).
The results are shown in Table \ref{tab_FLOP}.

The number of FLOP for our LBM implementation consists of two parts: computation of $v,\rho$ and collision.
The complexity of $v,\rho$ computations depends only on the fluid model.
These computations are implemented as a simple series of additions and optional divisions for the quasi-compressible model.
The computational complexity of collision depends practically almost only on collision model.
The full cost of the quasi-compressible model is then built-in into computations of $v,\rho$, where additional divisions are needed for each velocity component, and into the code responsible for boundary nodes handling (not shown in Table \ref{tab_FLOP}).
Due to a different ratio of division to collision complexity, the computational complexity of the quasi-compressible model implementation is from 50\% for LBGK to 14\% for LBMRT higher than for the incompressible model.

The comparison of FLOP/byte ratios from Table \ref{tab_FLOP} with specifications of high-performance GPUs (see Table \ref{tab_GPU}) shows that LBM implementations are usually bandwidth-bound.
This is especially true for the newest CUDA capable architectures, for which the FLOP/byte ratio usually gradually increases.
The high performance LBM implementations for GPU should then focus on memory bandwidth optimisations.
Notice that this is also true for general purpose CPUs, provided that their computational pow\-er is fully utilised using vector instructions.

\section{Implementation}
\label{sec_impl}

\subsection{Tiling overview}
\label{sec_tiling_overview}

\begin{figure}[!b]
\centering
\includegraphics{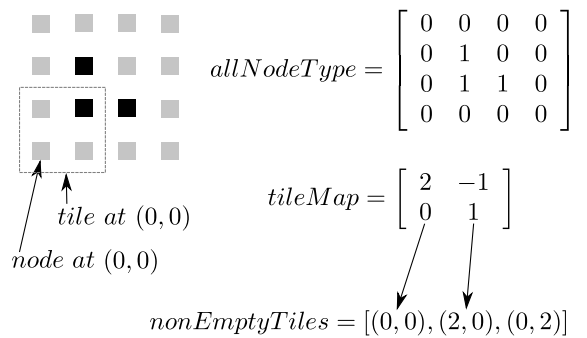}
\caption{
	Data structures used for tiling of 2D geometry containing $4 \times 4$ LBM nodes and tiles with $2^2$ nodes.
	Black/gray squares denote fluid/solid nodes.
}
\label{fig_tiling_data}
\end{figure}

In our implementation, the whole geometry is covered by the uniform mesh of cubic tiles, where each tile contains $a^3$ nodes.
If a geometry size is not divisible by $a$, then the geometry is extended with solid nodes.
The tiling is implemented by the host code (on CPU side) and done once at the geometry load.
We use a very simple tiling algorithm (shown in Algorithm \ref{alg_tiling}): first, the geometry is covered by the uniform mesh of tiles starting at node (0,0,0), and next, the tiles containing only solid nodes are removed.
The used data structures are shown in Fig. \ref{fig_tiling_data}.
The algorithm requires information whether each node in geometry is solid or not, thus we use \emph{allNodeType} array containing integers encoding the type of node (solid, fluid, boundary).

For each non-empty tile we store coordinates of its corner in one-dimensional \emph{nonEmptyTiles} array.
Coordinates of tile corner are placed in the order the non-empty tiles are found, thus we loose a connection between a geometric position of tile and index in \emph{nonEmptyTiles} array.
Since during LBM propagation the neighbour tiles must be located, we use an additional dense three-dimensional \emph{tileMap} matrix with indices to \emph{nonEmptyTiles} array (or -1 when the corresponding tile contains only solid nodes).
After tiling, the arrays \emph{nonEmptyTiles} and \emph{tileMap} are transferred to GPU memory.

The time and space complexities of the presented algorithm are linear as a function of the number of nodes (tile size can be treated as a constant).
For all geometries used in this work, the construction of presented data structures took less than a second on a single CPU core.

\begin{algorithm}[t]
\caption{
	General structure of the tiling algorithm for geometry containing $N_x \times N_y \times N_z$ nodes and tiles containing $a^3$ nodes.
	The algorithm requires input array containing node types for each node in geometry.
}
\label{alg_tiling}
\begin{algorithmic}[1]
	\For{$t_z = 0 ;~ t_z < N_z ; ~t_z += a$}
		\For{$t_y = 0 ;~ t_y < N_y ; ~t_y += a$}
			\For{$t_x = 0 ;~ t_x < N_x ; ~t_x += a$}
				
				\Statex
				\For{$n_z \in \left< 0..a-1 \right>$} \Comment{check nodes in tile}
					\For{$n_y \in \left< 0..a-1 \right>$}
						\For{$n_x \in \left< 0..a-1 \right>$}
							
							\State{$x = a \cdot t_x + n_x$}
							\State{$y = a \cdot t_y + n_y$}
							\State{$z = a \cdot t_z + n_z$}
				
							\If {node $(x,y,z)$ not solid}
								\State addNonEmptyTile $(t_x, t_y, t_z)$
								\State \textbf{goto} nextTile
							\EndIf

						\EndFor
					\EndFor
				\EndFor

				\Statex{nextTile:}
			\EndFor
		\EndFor
	\EndFor
\end{algorithmic}
\end{algorithm}

In general, the tile size could be arbitrary but, as we show below, it can have a significant impact on performance due to a low tile utilisation if a tile is too large.
Additionally, the number of nodes within a tile should be a multiple of the GPU warp size to avoid a low hardware utilisation.
In our implementation we have chosen $a = 4$.

\begin{figure}[tb]
\centering
\includegraphics{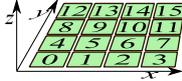}
\caption{
	Thread assignment to nodes inside a tile for \emph{threadIdx.z $= 0$}.
	Numbers denote linear thread index calculated as \emph{threadIdx.x + 4 $\cdot$ threadIdx.y + 16 $\cdot$ threadIdx.z}.
	Full thread block consists of 4 such planes along $z$ axis making 64 threads numbered from 0 to 63.
}
\label{fig_thread_numbering}
\end{figure}

Fig. \ref{fig_thread_numbering} shows thread assignment to nodes within a tile.
We use thread blocks with dimensions $\left< 4,4,t \cdot 4 \right>, t \in \left\{ 1, 2 \right\}$ so coordinates of nodes in the tile are simple to calculate: $x$ and $y$ node coordinates are the same as thread coordinates, $z$ coordinate of the node is equal to two the least significant bits of thread $z$ coordinate.
This mapping gives two warps per tile:
the first warp operates on all nodes with $z \in \{0,1\}$, the second warp on nodes with $z \in \{2,3\}$.

As in other GPU LBM implementations for sparse geometries \cite{Bernaschi:AFl2010, Nita:GPU2013, Huang:Mul2014}, we use two copies of $f_i$ values (denoted as $f_i$ and $f'_i$) to avoid race conditions.
Thus, the tiles can be processed independently and in any order during a single LBM iteration.
An additional research is required for the potential adaptation of techniques from \cite{Valero-Lara:Lev2016, Latt:How2007} that for dense geometries allow using only a single copy of $f_i$ functions.

Though the tiling is not a very novel technique (it is described even in NVIDIA tutorials as a method improving the performance of dense matrix multiplication, and was used in LBM implementation for IBM Cell processor presented in \cite{Sturmer:Flu2009}), to the best of our knowledge this is the first attempt to use it for an efficient GPU implementation of LBM for sparse geometries.
Existing GPU LBM implementations for sparse geometries \cite{Bernaschi:AFl2010, Nita:GPU2013,Huang:Mul2014} focus on indirect addressing, what does not guarantee proper memory layout and requires using additional indices that increase memory and bandwidth usage.
Below we will show that due to the regular memory layout the tiling offers 
a higher performance provided that the tiles contain a low number of solid nodes.

\subsection{Tiling memory layout}
\label{sec_tiling_memory_layout}

Since the LBM implementations are usually bandwidth-bound on high-performance GPUs, the proper memory layout resulting in good memory bandwidth utilisation is crucial to achieving high performance.
Thanks to the fixed tile size in our implementation we are able to fully control data placement for nodes processed in threads from the same warp.
This gives us a possibility to maximise the number of coalesced memory transactions, what results in a very high utilisation of GPU memory bandwidth (comparable to values reported for dense geometries).
The presented memory layout is optimised for double precision values.

\begin{figure}[tb]
\centering
\includegraphics{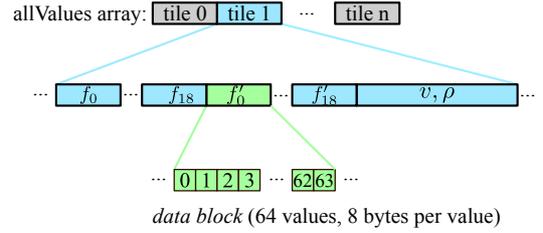}
\caption{
	Memory layout of array with all values ($f_i, v,\rho$) for nodes within tiles.
	We use two copies of $f_i$ values.
	For different $f_i$ we use different assignment of node coordinates to position in 64-element \emph{data block}.
}
\label{fig_general_memory_layout}
\end{figure}

The general memory layout used in our implementation is shown in Fig. \ref{fig_general_memory_layout}.
All floating point values for all nodes are stored in a single, continuous \emph{allValues} array, what allows us to control data placement in GPU memory.
In addition to the \emph{allValues} array, we also use additional data arrays, e.g.: \emph{nodeType} array where an information about LB nodes is stored, and \emph{tileMap} and \emph{nonEmptyTiles} arrays.
Data layout in the \emph{nodeType} array is similar to \emph{allValues} but for each tile, only one block of data is stored.
The \emph{tileMap} array is a simple three-dimensional array stored in row order.

\begin{figure}[!b]
\centering
\includegraphics{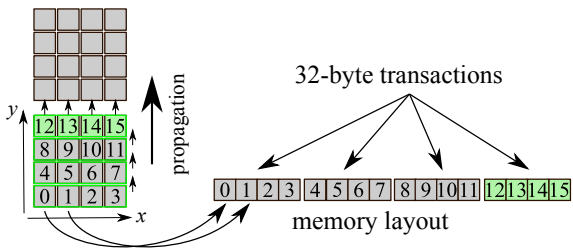}
\caption{
	Propagation in north (N) direction for XYZ memory layout and double precision $f_i$ values (8 bytes per value).
	Numbers denote linear memory offset of $f_i$ value for corresponding node (for example: $f_i$ for node at $x=0$, $y=1$ is placed at offset 4).
	Four consecutive double precision $f_i$ values (one row) form a single $32$-byte transaction.
	During propagation only four 32-byte memory transactions (reads) are required: one from the neighbour tile (read of nodes 12 to 15) and three from the current tile (read of nodes 0 to 11).
}
\label{fig_layout_XYZ}
\end{figure}

\begin{figure}[t]
\centering
\includegraphics{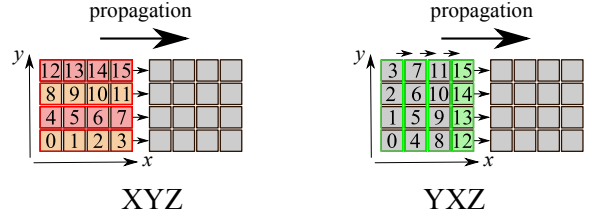}
\caption{
	Propagation in east (E) direction for two memory layouts and double precision $f_i$ values.
	For XYZ layout the four 32-byte wide memory transactions (marked with red rectangles) are needed to update nodes from the neighbour tile (nodes with indices 3,7,11,15).
	YXZ layout results in only one 32-byte transaction for nodes from the neighbour tile, because now these nodes have close indices (12,13,14,15).
Additionally, during the propagation inside the tile only three (instead of four) 32-byte read transactions are done for nodes with indices 0 to 11.
The fourth transaction is done from the neighbour tile.
}
\label{fig_layout_YXZ}
\end{figure}

\begin{figure}[!b]
\centering
\includegraphics{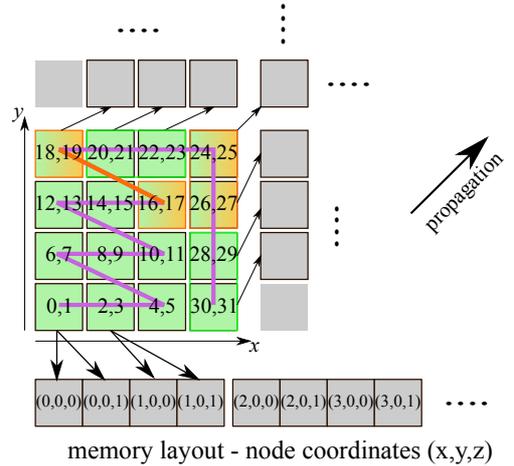}
\caption{
	Propagation in north-east (NE) direction for zigzagNE memory layout and double precision $f_i$ values.
	In this memory layout the two consecutive memory locations store $f_i$ values for nodes with the same $x$ and $y$ coordinates - only $z$ coordinate differs.
	Thus, each square on picture of tile denotes two 8-byte $f_i$ values placed in neighbour memory locations.
	Partially utilised (uncoalesced) memory transactions (offsets 16 to 19 and 24 to 27) are marked orange.
	Each orange transaction is done twice and only half of data is used.
}
\label{fig_layout_zigzag}
\end{figure}

A single value (e.g. $f_0$) from all nodes within the same tile forms a \emph{data block} shown in Fig. \ref{fig_general_memory_layout}.
Data layout in \emph{allValues} array guarantees that each data block (64 double precision values) is properly aligned and can be transferred using sixteen 32-byte wide memory transactions.
However, since during the propagation step some values must be read from nodes at edges of neighbour tiles, thus the standard row ordered data layout results in uncoalesced/redundant transactions (see for example Fig. \ref{fig_layout_YXZ}).
To address this issue, for each data block we applied an optimised data arrangement that allows reducing the number of memory transactions, which are not fully utilised (the orange/red transactions in Fig. \ref{fig_layout_YXZ}).
Each data arrangement is optimised for different access patterns during propagation (see Fig. \ref{fig_layout_XYZ}, \ref{fig_layout_YXZ} and \ref{fig_layout_zigzag}).

Let $x_t, y_t, z_t \in \{0,1,2,3\}$ denote node coordinates inside a tile.
In all below considerations we assume that $x_t, y_t, z_t$ are represented in the natural binary number system (unsigned types in CUDA/C language).
The transformation from the $x_t, y_t, z_t$ node coordinates to the offset inside \emph{data block} is defined by a linear mapping function $L (x_t, y_t, z_t)$.
We use three linear mapping functions (all equations are valid for $a=4$ nodes per tile edge):
\begin{equation}
L_{XYZ}(x_t, y_t, z_t) = x_t + 4 \cdot y_t + 4^2 \cdot z_t,
\label{equ_XYZ}
\end{equation}
\begin{equation}
L_{YXZ}(x_t, y_t, z_t) = y_t + 4 \cdot x_t + 4^2 \cdot z_t
\label{equ_YXZ}
\end{equation}
and
\begin{align}
L_{zigzagNE}(x_t, y_t, z_t) &= 2 \cdot \Big( x_t + 3 \cdot y_t +  \nonumber  \\
						 &+  \big( (x_t + 1) \cap 4 \big) \cdot (3 - y_t) \Big)  +  \\
						 &+ (z_t \cap 1)  + 4^2 \cdot (z_t \cap 2) \nonumber
,
\label{equ_zigzagNE}
\end{align}
where $\cap$ denotes bitwise AND.

All mapping functions allow to minimise the number of uncoalesced memory transaction for our thread mapping and tile size.
In our implementation we have started from the $L_{XYZ}$ layout for all $f_i$ data blocks and, after code profiling, we have modified layouts for these $f_i$ data blocks, for which the redundant memory traffic was observed.
$L_{XYZ}$ have been used for $f_{O}, f_{N}, f_S, f_T, f_B, f_{NT}, f_{NB}, f_{ST}, f_{SB}$, $L_{YXZ}$ have been used for $f_E$, $f_W$, $f_{ET}$, $f_{EB}$, $f_{NW}$, $f_{SW}$, $f_{WT}$, $f_{WB}$
and $L_{zigzagNE}$ have been used for $f_{NE}$ and $f_{SE}$.
For this mapping functions assignment only for four $f_i$ functions ($f_{NE}, f_{SE}, f_{NW}, f_{SW}$) the number of memory transactions is not minimal ($64 ~ \lbrack nodes \rbrack ~ \cdot ~ 8 ~\lbrack bytes \rbrack~ /$ $ 32 ~\lbrack bytes/trans. \rbrack$ $= 16$ transactions per $f_i$ per tile).
Both $f_{NE}$ and $f_{SE}$ require four additional memory transactions per tile (two per warp - marked orange in Fig. \ref{fig_layout_zigzag}).
Propagation of $f_{NW}$ and $f_{SW}$ requires 32 transactions per tile each, what gives 100\% overhead.
For $f_{NW}$ and $f_{SW}$ we also tried to use a mapping function similar to $L_{zigzagNE}$, but it resulted in slightly decreased performance despite the lowered number of memory transactions.

The total number of memory transactions required for propagation of a tile is then
$
15 ~\lbrack f_i~values \rbrack \cdot 16 ~\lbrack transactions \rbrack $ $
 +~2~\lbrack f_i~values \rbrack \cdot (16+4)~\lbrack transactions \rbrack
+ 2~\lbrack f_i~values \rbrack \cdot 32 ~\lbrack transactions \rbrack
= 344
$
transactions. 
This gives a 13\% overhead compared with the minimal number of transactions ($19 \cdot 16 = 304$).

\subsubsection{Single precision overheads}
\label{sec_single_prec_transactions}

For single precision values, the presented data layout also allows to reduce the number of memory transactions, but the significantly larger overhead is observed.
The main reason is that only two 32-byte transactions are needed for one layer of $f_i$ values (16 4-byte numbers), thus it is impossible to completely utilise full memory transaction while reading only one row of values from a neighbour tile.

The minimum number of transactions per single $f_i$ function is 8, what gives $19 \cdot 8 = 152$ transaction per complete tile.
For XYZ layout this minimal value can be preserved only for three functions - $f_O, f_T$ and $f_B$.
Six functions ($f_N, f_S, f_{NT}, f_{NB}, f_{ST}, f_{SB}$) require 12 transactions per $f_i$, because for each layer of $f_i$ values the two transactions from current tile and one transaction from neighbour tile are needed (use in Fig. \ref{fig_layout_XYZ} transactions containing two rows instead of one).
Next six functions ($f_E$, $f_W$, $f_{ET}$, $f_{EB}$, $f_{WT}$, $f_{WB}$) require 16 transactions per $f_i$, similar to Fig. \ref{fig_layout_YXZ} left.
The last four functions, $f_{NE}, f_{SE}, f_{NW}, f_{SW}$, use 24 transactions per $f_i$.
The total number of memory transactions for single precision values is then 288, what gives almost 90\% overhead compared to the minimal value.

After using the presented memory layout the numbers of transactions per eight functions ($f_E$, $f_W$, $f_{ET}$, $f_{EB}$, $f_{WT}$, $f_{WB}$, $f_{NE}, f_{SE}$) reduce to 12 transactions per $f_i$
This allows gathering all $f_i$ values during propagation step using 240 memory transactions.
Though this value is 17\% smaller than for XYZ layout, it still results in 58\% overhead compared to the minimal 152 transactions per tile.

\subsection{Tiling overhead}
\label{sec_tiling_overhead}

Since tiles have cubic shape and the fixed size, we can minimise the number of uncoalesced memory transactions per tile, as shown in Sec. \ref{sec_tiling_memory_layout}.
Unfortunately, we always must use full tiles.
For this reason, the tile based implementation may need to store data and do computations not only for fluid nodes but also for some neighbour, solid nodes that fall into the tiles with non-solid nodes (see Fig. \ref{fig_tiling_square_8}).
Compared with the minimal requirements defined in Sec. \ref{sect_performance_model}, the tiling brings some memory, bandwidth and computational overhead.
The three main reasons for the tiling overhead are additional solid nodes inside tiles, two copies of $f_i$ values, and some other data (e.g. $tileMap$ array).
Furthermore, the data placement can cause an additional bandwidth overhead as shown in Sec. \ref{sec_tiling_memory_layout}.

Let the overhead $\Delta$ be defined as a ratio of additional memory/operations ($\Delta^M$ - memory overhead, $\Delta^B$ - bandwidth overhead, $\Delta^C$ - computational overhead) to the minimum numbers defined in Sec. \ref{sect_performance_model}.
Only one LBM time iteration is analysed since all iterations are processed in the same way.

Let $a$ denote the number of nodes per tile edge, $n_{tn} = a ^ 3$ denote the number of nodes per tile, $t_n$ denote the number of non-empty tiles (tiles with at least one non-solid node), and $n_{fn}$ denote the number of non-solid nodes in the whole geometry.
Since we process only non-empty tiles, then the overheads depend on the ratio of $n_{fn}$ to the number of all nodes in all non-empty tiles.
For brevity, we define the average tile utilisation factor
\begin{equation}
	\eta_t = \frac{n_{fn}} {t_n \cdot n_{tn}}
	,
	\label{equ_tileutilfac}
\end{equation}
the average number of non-solid nodes per tile $n_{tfn} = n_{fn} / t_n$, and the average number of solid nodes per tile $n_{tsn} = (t_n \cdot n_{tn} - n_{fn}) / t_n = n_{tn} - n_{tfn}$.
Notice that $\eta_t = n_{tfn} / n_{tn}$.

When we assume that for each node the number of operations and used memory are minimal (i.e. that the tiling does not have an impact on the computational complexity for single nodes), then the overhead caused by low $\eta_t$ results only from additional operations/memory, which must be done/allocated for all solid nodes from tile.
In this case, the generic overhead value is then
\begin{equation}
	\Delta_{\eta_t} = \frac{n_{tsn}}{n_{tfn}} = \frac{n_{tsn}}{n_{tn} - n_{tsn}} = \frac{1 - \eta_t}{\eta_t}
	.
	\label{equ_generic_delta}
\end{equation}
In our implementation, the real values of bandwidth and computational overheads are even lower because we conditionally skip operations for solid nodes.
For computational overhead, some operations are not done when the full warp of threads processes only solid nodes.
This situation is rather rare (but possible), as a half of a tile must contain solid nodes.
For bandwidth overhead, the transfers are omitted when values for solid nodes form a single 32-byte memory transaction.
This happens quite often due to a spatial locality of geometry.
Also, some transfers can be skipped, when during the propagation step the neighbour tile contains solid nodes.

For the memory overhead,
Eqn. (\ref{equ_generic_delta}) 
allows only to find the overhead compared to an implementation where the amount of data stored per node is the same as in the tile-based implementation.
However, if we use as a reference the value defined by Eqn. (\ref{equ_mnode}), then the overhead must take into account the second copy of $f_i$ and additional $n_t$ bytes used in our implementation to encode the type of node (solid, fluid, boundary, thus usually $n_t \in \left<1,4 \right>$)
\begin{align}
	\Delta^M_{\eta_t} & = \frac {n_{tn} \cdot (2 \cdot q \cdot n_d + n_t) - n_{tfn} \cdot q \cdot n_d}
															{n_{tfn} \cdot q \cdot n_d} = \nonumber \\
					 & = \frac {2 \cdot q \cdot n_d + n_t} {\eta_t \cdot q \cdot n_d} - 1 
					 \approx \frac{2 - \eta_t}{\eta_t}
.
\label{equ_delta_m_single}
\end{align}
The final approximation is true for $n_t \ll q \cdot n_d$.
Hence, the memory overhead compared with minimal requirements defined by Eqn. (\ref{equ_mnode}) grows about twice as fast as the overhead defined by Eqn. (\ref{equ_generic_delta}).

In addition to Eqn. (\ref{equ_generic_delta}) and (\ref{equ_delta_m_single}), the memory and bandwidth overheads are also affected by the $tileMap$ and $nonEmptyTiles$ arrays, but they usually may be skipped.
For each non-empty tile, $3^3$ values from the $tileMap$ array (neighbour tiles) and a coordinates of current tile  from $nonEmptyTiles$ are read, and for each tile (both empty and non-empty), a single value must be stored in $tileMap$.
Each value in $tileMap$ can be at most 4-bytes long because $2^{32}$ tiles require much more memory, than is available in current GPUs (for D3Q19 lattice, double precision data and tile containing $4^3$ nodes the data for all nodes require about $2^{32} \cdot 4^3 \cdot M_{node} \approx 39$ TiB of memory).
Coordinates of tile can be also packed into a 4-byte number, for example by using functions similar to defined in Eqn. (\ref{equ_XYZ}).
Single tile requires $ 4^3 \cdot M_{node} \approx 9800$ bytes, what is almost 2500 times larger than the index in $tileMap$.
Thus, except for exceptionally sparse geometries (consisting significantly less than 1\% of non-empty tiles), the overhead caused by $tileMap$ and $nonEmptyTiles$ arrays is negligible.

It can be seen from Eqn. (\ref{equ_generic_delta}) and (\ref{equ_delta_m_single}) that the overhead grows quickly when the tile utilisation factor is lower than 1
(the overhead is 0.2 already for $\eta_t = 0.83$).
Therefore, it is critical to achieve the tile utilisation as high as possible.

\begin{figure}[!t]
\centering{
\subfloat{\includegraphics{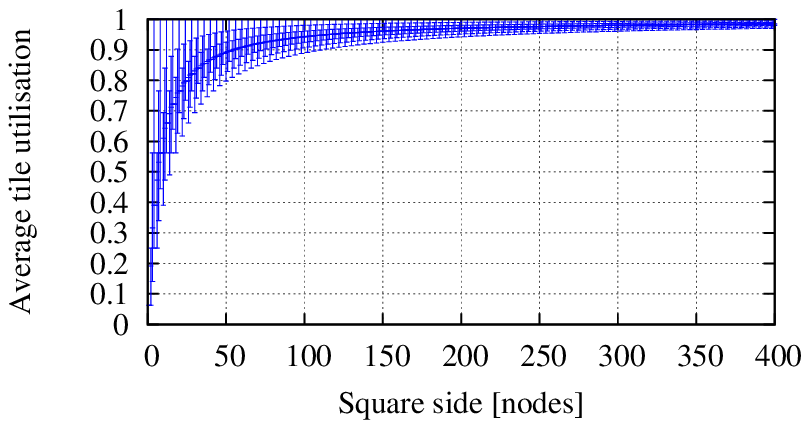}}
\hfill
\subfloat{\includegraphics{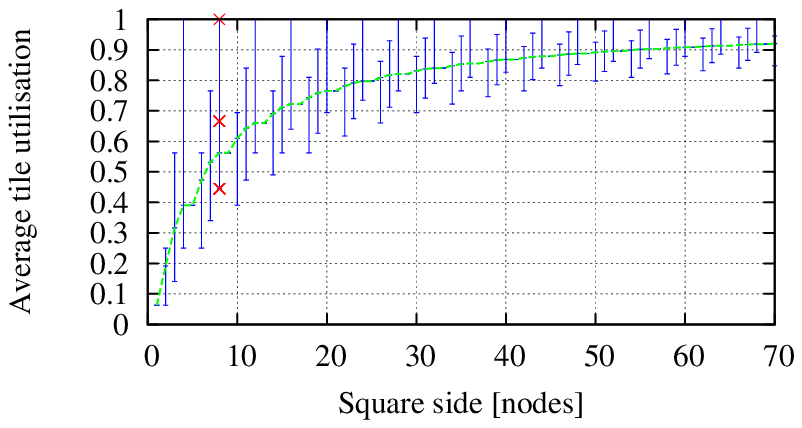}}
\hfill
\subfloat{\includegraphics{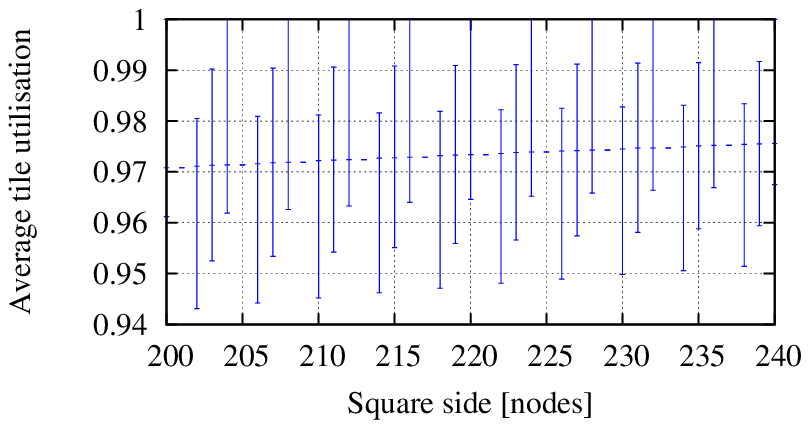}}
}
\caption{
Average tile utilization for all available tilings of infinitely long square channel along the axis.
For each channel size sixteen tilings are possible (see Fig. \ref{fig_tiling_square_8}).
Green line denotes the average value of average tile utilisation for given channel dimension (see Fig. \ref{fig_tiling_square_8}).
Red crosses denote real tile utilization for channel $8 \times 8$ nodes - there are only 3 available values.
}
\label{fig_tile_util_square}
\end{figure}

\begin{figure}[thbp]
\centering
\includegraphics{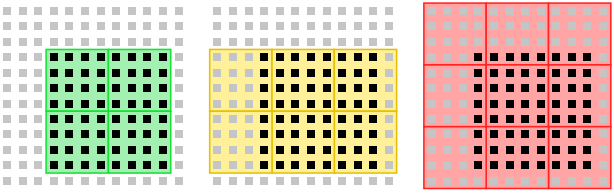}
\caption{
Cross section of 3 example tilings (marked on Fig. \ref{fig_tile_util_square}) of a square channel $8 \times 8$ nodes.
Black/grey squares denote fluid/solid nodes.
The channel is parallel to one of the axes.
Average tile utilisation is $64/(4\cdot16) = 1.0$, $64/(6 \cdot 16) \approx 0.67$ and $64 / (9\cdot 64) \approx 0.44$ for green, yellow and red tilling.
For this channel and tile size there are 9 tilings with average tile utilisation $0.44$, 6 tilings similar to yellow tiling and only one green tiling with tile utilisation equal to 1.
The average value of average tile utilisation is then $(9 \cdot 0.44 + 6 \cdot 0.67 + 1)/16 \approx 0.56$.
}
\label{fig_tiling_square_8}
\end{figure}

\begin{figure}[!t]
\centering{
\subfloat{\includegraphics{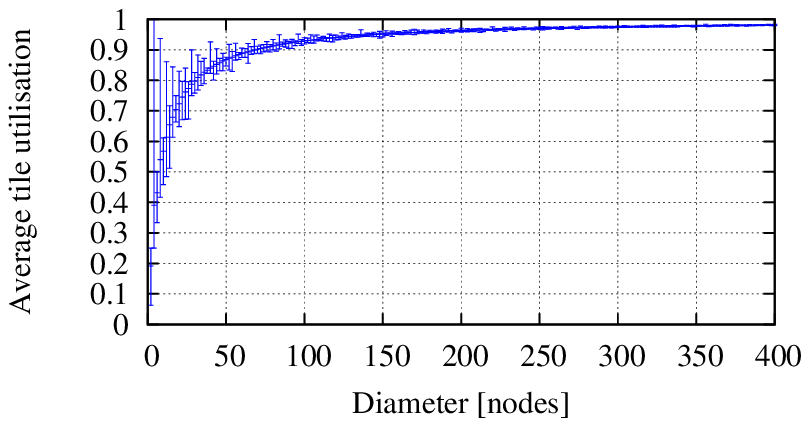}}
\hfill
\subfloat{\includegraphics{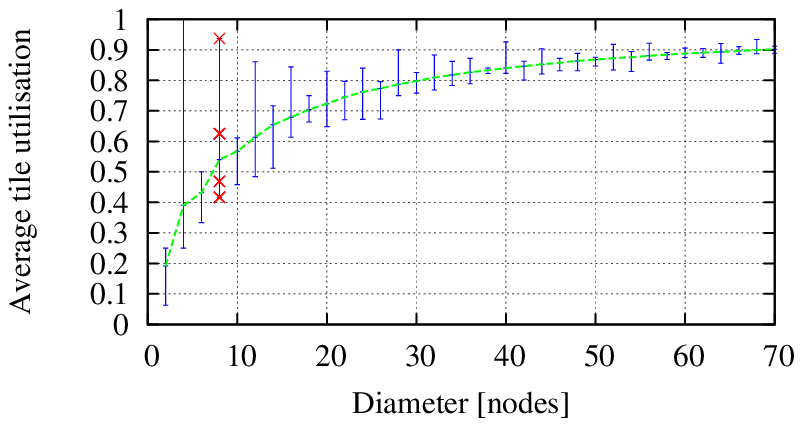}}
}
\caption{
	Average tile utilisation for all available tilings of infinitely long circular channel along the axis (16 different tilings are possible for each diameter).
	Green line denotes the average value of average tile utilisation for given channel size (see Fig. \ref{fig_tiling_square_8}).
	Red crosses denote real tile utilization for channel with 8 nodes diameter - in this case only four possible values can occur.
}
\label{fig_tile_util_circle}
\end{figure}

In Fig. \ref{fig_tile_util_square} and \ref{fig_tile_util_circle} we show, how $\eta_t$ changes for tiling of infinitely long channels running along one of the axes.
Notice that the tile positions are discrete, thus only a few values of $\eta_t$ can occur.
It can be seen that tile utilisation above $0.8$ can be always achieved for channels with diameter/side at least about 40 nodes.
To have a guarantee that the tile utilisation is always above $0.9$, the channels must have about 100 nodes across.
Though these channel dimensions can be too large for some applications, it should be noted that these values are pessimistic.
It is possible to find a tile placement that gives good tile utilisation even for very small channels (for example, $\eta_t$ can be equal to 1 for a square channel as small as $4\times 4$ nodes).

Additionally, since channels often are not parallel to the axes, then the values of $\eta_t$ for real cases can be closer to the average value for different tilings of the same channel size (see Fig. \ref{fig_tiling_square_8}).
The average value for different tilings of the same channel size (the green line in Fig. \ref{fig_tile_util_square} and \ref{fig_tile_util_circle}) exceeds $\eta_t = 0.8$ for square channel $25^2$ nodes and circular channel with diameter 30 nodes.
For channel dimensions about 55 and 65 nodes for square and circular channels $\eta_t$ exceeds $0.9$.

Some observations can be also extracted from Fig. \ref{fig_tile_util_square} and \ref{fig_tile_util_circle}.
It can be seen that 
the dispersion of the tile utilisation for different tilings of a channel with given dimension is higher for square channels than for circular channels.
It is possible to achieve higher tile utilisation for square channels, but it is also possible that the tile utilisation will be much lower than for circular ones.

The average value of different tile utilisations is higher for circular channels.
The difference between the tile utilisation for circular and square channels depends greatly on the channel dimension: starting from 3 times for the smallest channels, through the 10\% for channels with 20 nodes at diameter/side, and to the less than 2\% for channels with at least 100 nodes.

For square channels, the two additional phenomena occur.
First, if the channel dimension is larger by one node than the tile edge, then all tilings have the same tile utilisation.
In this case, all tilings require the same number of tiles.
The one node outside tile borders can always be placed in either "left" or "right" side of the tile, never on both sides.
When the channel dimension is larger by two or more nodes than the tile edge, the number of additional tiles may differ depending on the tile placement.
For example, two additional nodes can be put either in one additional tile or in two additional tiles on both sides.
This behaviour can be used during geometry preparation to guarantee constant tile utilisation.

Second, the tile utilisation fluctuates with a period equal to the tile edge $a = 4$ nodes, i.e. for channel edge equal to $k \cdot a + m$, where $m \in \lbrace 0,1, \ldots, a-1 \rbrace$ and integer $k \geq 0$.
This results from the fact that for a single period (for the same $k$) the same numbers of tiles are needed: $k$, $k+1$ or $k+2$ tiles along each axis.
Since the number of non-solid nodes grows with $m$, then for the same number of tiles the tile utilisation $\eta_t$ becomes larger.
As the lowest tile utilisation is always for $m=2$, square channel side should be different from $k \cdot 4 + 2.$

\subsection{Implementation details}

Our code is written in CUDA C language ver. 7.5, the first version with official support for many C++11 features.
Thanks to this we could write a highly templated code, which can be shared between the CPU and GPU compilers.
The code sharing enabled us to simplify testing because results of many parts of the code could be thoroughly checked only on the CPU side.
Additionally, extensive use of C++ templates allowed us to write a generic kernel code, which later can be specialised for fluid and collision models, data layouts, data precision, enabled optimisations etc.
We were then able to test many combinations of the above parameters with minimal changes to source code and without code duplication.
The disadvantages of this solution are a long compilation time and a large size of the executable file containing many versions of the same kernel.

In our implementation, the computations for a single LBM time step include collision, propagation and boundary computations for all nodes within the tiles. 
To minimise memory transfers, we combine collision, propagation and boundary computations for a single node into a single GPU kernel.
The kernel requires only one read of data from memory at the beginning, and one store of data at the kernel end.
Local copies of data are stored in registers and in the shared memory.
Separate tiles are processed by a separate GPU thread blocks (see Fig. \ref{fig_thread_numbering}), what allows for effortless synchronisation of computations within the tile.
The general structure of our GPU kernel is shown in Algorithm \ref{fig_kernel}.

\begin{algorithm}[thb]
\begin{algorithmic}[1]
	\State \textbf{load} copy of $tileMap$ \Comment $3^3$ values
	\State \textbf{load} node types from current tile \Comment $4^3$ values
	\State \textbf{load} node types from neighbour tiles \Comment WLP
	\State \textbf{BARRIER}
	\If {node not solid}
		\State \textbf{load} $f_0$ 
		\For{$q \in \left< 1..18 \right>$}
        \State \textbf{compute address} of neighbour node in direction $q$
				\If {neighbour node not solid}
					\State \textbf{gather} $f_q$ from neighbour node
				\EndIf
    \EndFor \Comment all $f_i$ copied to registers
		\State process boundary \Comment{calculate $v,\rho$}
		\State collide
		\State \textbf{store} all $f_i$ \Comment all coalesced
	\EndIf
\end{algorithmic}
\caption{Structure of the GPU kernel implementing single LBM time iteration for a single node.}
\label{fig_kernel}
\end{algorithm}

The propagation is implemented as a gathering data from neighbour nodes.
The propagation between two nodes is done only when the target and source nodes are not solid.
To minimise a cost of checking node types, we use shared memory to store copies of node types from current and neighbour tiles (lines 2--3 in Algorithm \ref{fig_kernel}). 
In code responsible for loading node types from neighbour tiles (line 3) we use warp level programming to balance the load of warps assigned to the tile.
We also make a copy of $3^3$ tile indices from the $tileMap$ array forming a cube surrounding currently processed tile (line 1).
All later operations requiring information about node and/or tile types use these copies.

\begin{figure}[b]
\centering
\includegraphics{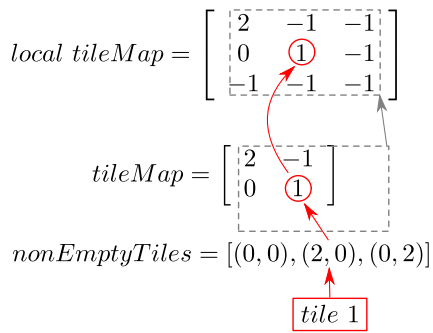}
\caption{
	Steps required for making a local copy of $tileMap$ array for 2D geometries and tiles containing $2^2$ nodes.
	The geometry is identical as in Fig. \ref{fig_tiling_data}.
}
\label{fig_tileMap}
\end{figure}

The operations required to make a local copy of $tileMap$ are shown in Fig. \ref{fig_tileMap}.
For simplification, we reduced the geometry to two-dimensional.
The local copy is stored in shared memory and used by all threads processing the tile.
First, the tile index is computed based on the index of current CUDA block (red rectangle in Fig. \ref{fig_tileMap}).
When the tile index is known, the coordinates of tile corner can be loaded from $nonEmptyTiles$ array.
Then the corner coordinates are transformed into indices of an element in the global $tileMap$ array (the reverse operation to presented in Fig. \ref{fig_tiling_data}).
In the last step, the local copy is filled with the 
fragment of $tileMap$ array containing $3 \times 3$ elements and
centred at the element with tile corner coordinates.
The resulting local copy of $tileMap$ contains indices to all tiles surrounding the currently processed one.

\begin{figure}
\centering
\includegraphics{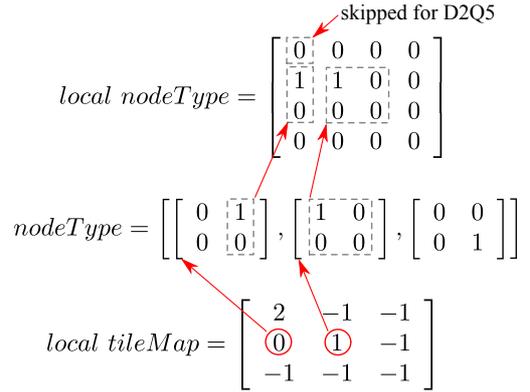}
\caption{
	Steps required for making a local copy of $nodeType$ array for D2Q5 lattice arrangement and tiles containing $2^2$ nodes.
	The geometry is identical as in Fig. \ref{fig_tiling_data}.
}
\label{fig_nodeType}
\end{figure}

The process of making local copy of $nodeType$ array is presented in Fig. \ref{fig_nodeType}.
Since the indices of all neighbour tiles are already cached in the local copy, the only operations on global memory use the $nodeType$ array containing the node type fields for nodes from non-empty tiles.
Two steps are needed for each neighbour tile: first, the address of data block containing node types for the tile is computed, then the required node type values are copied to shared memory.
Notice that the local copy is always 2 nodes per side larger than the tile to allow for storing node types from neighbour tiles.
Also, only these neighbour tiles are visited that result from lattice linkage. 

After loading of node and tile types, the barrier is needed (line 4 in Algorithm \ref{fig_kernel}) because in a later part of the kernel the threads from different warps require the data gathered in other warps.
This is the only synchronisation point in the kernel.

All later operations are done only when a node assigned to current thread is not solid (line 5).
In this way, we can skip unnecessary operations as mentioned in Sec. \ref{sec_tiling_overhead}.

The computations for a non-solid node are divided into the three parts: lines 6--12 are responsible for gathering $f_i$ values, lines 13 and 14 perform all floating point calculations and line 15 stores the computed $f_i$ values to the memory.
All operations are implemented as standard except for
the line 8, where we compute indices for a neighbour node.
First, the index of a tile containing the neighbour node is computed - we use only the values from $\left< -1,0,1 \right>$ due to the local copy of the $tileMap$ array.
When the neighbour tile is not empty, a few indices for the neighbour node are computed since we need both information about the neighbour node type (this is gathered from the local copy of $nodeType$ array stored in shared memory) and the value of proper $f_i$ (which is stored in global memory in $allValues$ array).
Notice that for different $f_i$ functions we need to use the different data layouts (see Sec. \ref{sec_tiling_memory_layout}).

In the GPU kernel, we have also applied some optimisations not shown in Algorithm \ref{fig_kernel}.
First, the loop starting at line 7 is partially unrolled to allow computations for two $f_i$ functions in a single iteration, what increases the instruction level parallelism due to the interlacing of two independent streams of instructions.
In addition, the order of iterations over $q$ is chosen in a way that allows sharing some parts of neighbour node address computations (line 8).
Next, some values (e.g. $v$) are stored in shared memory to minimise register pressure.
Shared memory is also used in the implementation of the bounce back boundary condition to avoid a significant increase of required registers.
To minimise conflicts in shared memory, 64-bit mode is used.
We have also tuned the usage of registers for each specialisation of the kernel since different collision models require different amounts of temporary data.
The cost of divergence in the code responsible for computations of an address of the neighbour node is minimised - only integer indices are computed in divergent branches, time-consuming operations like memory accesses are done in all threads simultaneously.
In places, where it is possible and useful, the number of costly divisions is reduced by replacing them with multiplications by the inverse.
Finally, we have been intensively using inline methods and compiler pragmas to force loop unrolling, what allowed us to connect clean, structured code with high performance.

\section{Results}
\label{sec_results}

\subsection{Measurements methodology}

All tests have been run on a computer with the Intel i7-3930K CPU with 64 GB quad-channel DDR3 DRAM and GTX Titan GPU (Kepler architecture) clocked at 823 MHz with 6 GB GDDR5 memory clocked at 3.004 GHz.
The peak theoretical GPU memory bandwidth is $288.384$ GB/s (GB/s denotes $10^9$ bytes/s), the NVIDIA \emph{bandwidthTest} utility reported $228.5$ GB/s (79.2\% of peak).
We have been using Linux operating system with NVIDIA CUDA Toolkit 7.5 installed.

To identify performance limits we have been using three versions of kernels: 
kernels implementing all LBM operations, 
kernel with full propagation but without computations,
and kernel without propagation, which only reads and writes node data for the same node (without communication between neighbour nodes).
The kernels implementing all LBM operations are described by collision and fluid models (e.g. "LBGK incompressible"), the kernel with propagation is marked as "propagation only" and the last kernel without full propagation is denoted as "read/write only".
All kernels were prepared in two versions operating on double and single precision floating point numbers.

As a measure of kernel performance, we have been using the number of $10^6 \times$ non-solid (fluid and boundary) nodes processed per second (MFLUPS) \cite{Valero-Lara:Lev2016}.
The kernel performance was measured on the final version of the code with the main computational loop containing not only the launches of optimised kernel, but also the conditional stores of results to disk and the launches of the modified kernels used to estimate the convergence.
Since the impact of the initial data preparation on the simulation performance strongly depends on case configuration (for example, on the number of required iterations), we focused only on the performance of computational kernel repeated iteratively.
Moreover, for cases used in the paper the full simulation requires up to more than $10^5$ LBM iterations, then the initial data preparation can be safely skipped.
The same applies to additional operations inside the computational loop (data writes to disk or diagnostic messages) that are costly compared to single kernel run and should be done only when necessary.
Thus, to take into account only the time spent in the profiled kernel we measured the wall clock time for each kernel launch separately and with necessary synchronisation at kernel end.
For time measurement we have been using the C++ \emph{std::chrono} library with microsecond resolution.

\begin{figure}[htbp]
\centering{
\subfloat{\includegraphics{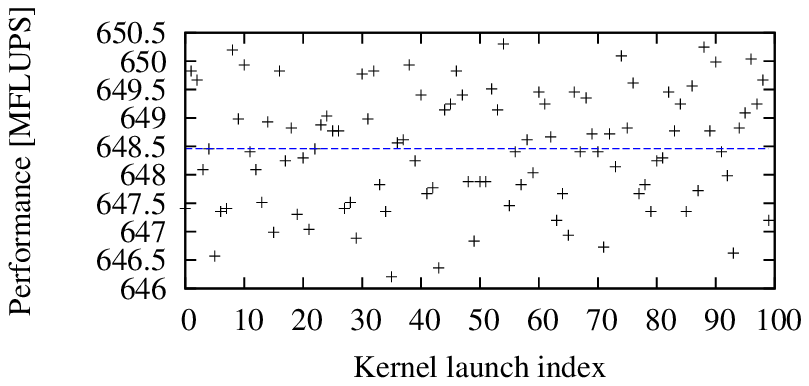}}
\hfill
\subfloat{\includegraphics{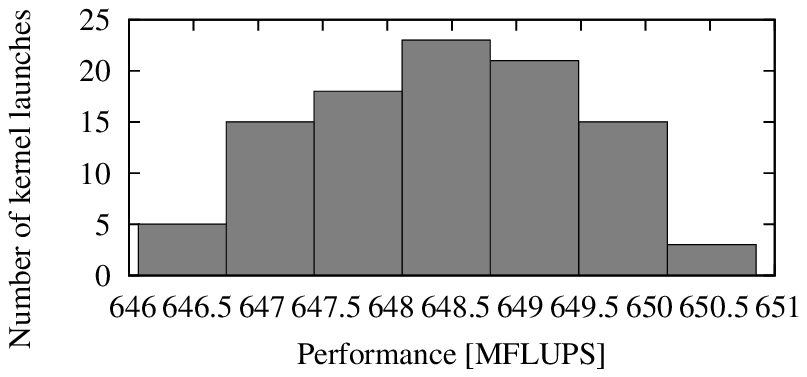}}
}
\caption{
	Performance of subsequent kernel launches (top) and histogram of kernel performance (bottom) for double precision LBGK incompressible flow simulation in cavity3D geometry containing $200^3$ nodes.
	The blue dashed line denotes the average kernel performance.
}
\label{fig_measurements_200}
\end{figure}

The overhead of this method was negligible in our system - the difference between the sum of measured separate stages of the computational loop and the total wall clock time spent in the whole computational loop was less than one microsecond per single kernel launch.
The single kernel took from about 30 microseconds (for $20^3$ nodes) up to tens of milliseconds (for the largest geometries).
Each version of the optimised kernel was run 100 times in succession.
Sample results are shown in Fig. \ref{fig_measurements_200}.
The observed difference between the shortest and the longest measured time spent in separate kernel launches was 78 $\mu$s for data from Fig. \ref{fig_measurements_200}.
For the smallest geometries, the difference was less than 15 $\mu$s.

\subsection{Thread mapping}

The thread mapping described in Sec. \ref{sec_tiling_overview} allows assigning an arbitrary number of tiles per thread block.
This enables to increase the GPU occupancy on machines, where the maximum number of simultaneously residing blocks is small.
For example, GK110 GPU allows for up 16 blocks residing on a multiprocessor, what for 64 threads per block (single tile contains 64 nodes) results in 50\% occupancy.
However, the increase of occupancy by assigning more threads per block is possible only when threads use a small number of registers.
Since our double precision kernels require at least 64 registers per thread, the maximum occupancy can not exceed 50\% for GK110 GPU.
Thus, we used larger blocks (containing 128 threads that process two tiles) only for single precision kernels, which achieved the highest performance for 40 or 48 registers per thread.
The disadvantage of the larger blocks is more complex computations in the kernel because calculations of indices for internal data structures are slightly more complex than for single tile per block.

\subsection{Performance for cavity3D}

\begin{figure}[t]
\centering{
\subfloat{\includegraphics{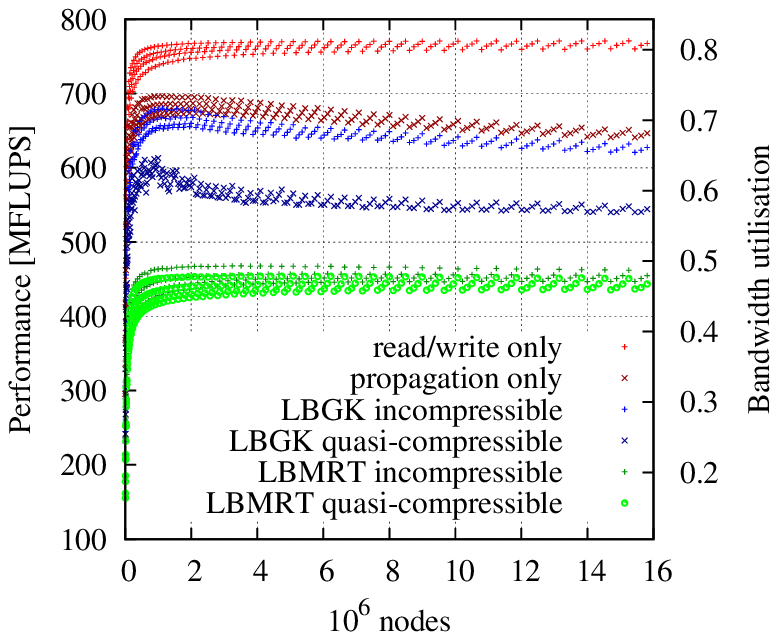}}
\hfill
\subfloat{\includegraphics{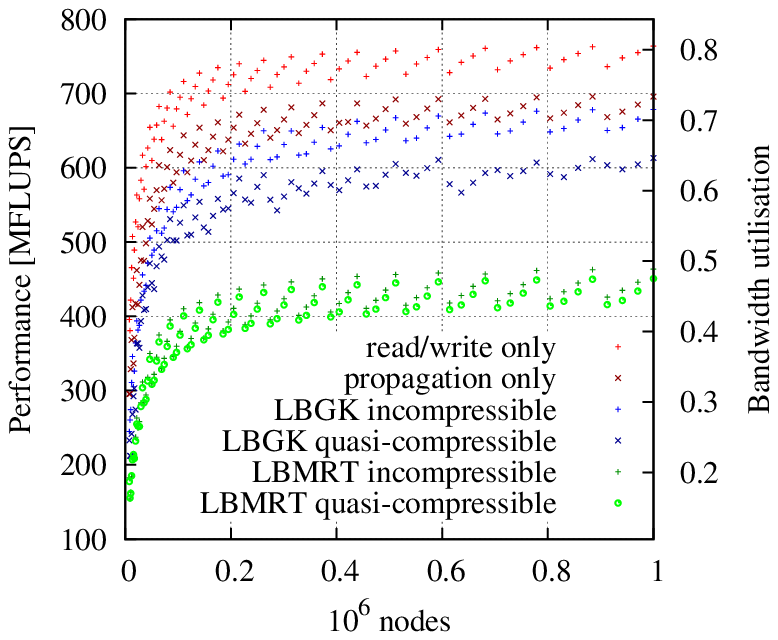}}
}
\caption{
Kernel performance for double precision kernels and cavity3D with geometry size $b^3$ nodes, where $b \in \left< 20, 252 \right>$. 
All measures on GTX Titan.
}
\label{fig_microflow_mlups_cav3D_double}
\end{figure}

To identify performance limits not affected by the geometry sparsity, we have first measured the performance of our implementation for the cavity3D test case, a standard example of dense geometry.
The performance of all kernels as a function of a geometry size is shown in Fig. \ref{fig_microflow_mlups_cav3D_double}.

The performance dependency on the geometry size is similar to many other GPU implementations.
For $10^5$ nodes the kernels achieve about 85\% of their maximum performance.
To achieve more than 95\% of maximum performance, at least $3 \times 10^5$ nodes are needed.
For other GPUs, the numbers of nodes can be different, but for today's machines, the maximum performance should always be possible to achieve for geometries containing about $10^6$ or more nodes.

The highest performance was observed for read/write only kernel: 771 MFLUPS what, according to Eqn. (\ref{equ_bnode}), corresponds to memory bandwidth equal to $234.4$ GB/s (81.3\% of maximal theoretical bandwidth for GTX Titan).
Because in this kernel all memory transactions are coalesced, we achieved bandwidth utilisation even higher than reported by the $bandwidthTest$ utility.

For kernel with propagation only, the performance drop\-ped by 9.7 \% to 696 MFLUPS resulting in bandwidth 211.6 GB/s (73.4\% of maximum).
It should be noted that though the propagation only kernel has more than seven times more instructions than read/write only kernel (711 vs 99 instructions),  the performance penalty caused by neighbour tile data access during propagation is smaller than 10\%.
More than 60\% of all instructions in propagation only kernel are integer arithmetic operations used for address computations.

Finally, after enabling all computations the kernel performance dropped by an additional 2.3\% for LBGK incompressible (680 MFLUPS, 71.7\% of GPU memory bandwidth) up to 34.8\% for LBMRT quasi-compressible (454 MFLUPS, 47.9\% of GPU memory bandwidth).
The large performance drop for LBMRT results from computational overhead - due to the large register usage per single thread and the low GPU occupancy the complex double precision computations are not masked with memory transfers.

The performance of our implementation for sparse geometries is close even to the highly optimised version for dense geometries from \cite{Mawson:Mem2014}, which utilises $76.7$\% of the maximum memory bandwidth for Tesla K20c and the LBGK quasi-compressible model.
Moreover, the implementation from \cite{Mawson:Mem2014} uses single precision numbers what allows to increase GPU utilisation due to much lower register pressure.
This result shows that it is possible to process sparse geometries with performance similar to dense ones.

As can be seen in Fig. \ref{fig_microflow_mlups_cav3D_double}, the performance of all kernel versions fluctuates for geometry edge $b = 4 \cdot k + m$, where $m \in \lbrace 0, \ldots, 3 \rbrace$.
This results from the low tile utilisation for tiles placed on the geometry edges (see Section \ref{sec_tiling_overhead}).
Notice also that a performance difference for the geometry edge $b$ computed for the same $k$ and different $m$ decreases for larger $k$, e.g. for $k = 25 \rightarrow b \in \{100,101,102,103 \}$ the performance of propagation only kernel is between 677 and 696 MFLUPS, whereas for $b \in \{248, 249, 250, 251 \}$ between 641 and 651 MFLUPS.
This results primarily from differences in the average tile utilisation caused by partially filled tiles at geometry boundaries for $m \neq 0$.
For the cubic cavity3D geometry the number of fully utilised tiles inside the geometry grows proportionally to $b^3$ and the number of partially utilised tiles (at boundary walls) proportionally to $b^2$.
This causes the growth of the average tile utilisation for larger $b$.

Some interesting behaviour is also shown in Fig. \ref{fig_microflow_mlups_cav3D_double} - after achieving peak about $10^6$ nodes the performance slowly decreases with a geometry size increase.
The intensity of this phenomenon depends on a computational complexity of operation - the most visible is for the propagation only kernel (see Table \ref{tab_perf_spread}).
Also, it does not occur for the kernel without propagation (read/write only).
This is a result of data propagation between neighbour tiles that increases the time needed for tile processing -- for read/write only kernel there is no propagation between tiles, thus no performance decrease is observed.
For smaller geometries the ratio of tile faces and edges common to two tiles to the number of tiles is smaller than for large geometries, thus the performance for smaller geometries is higher.

\begin{table}
\caption{
	Kernel performance in MFLUPS for dense geometry cavity3D and double precision kernels. 
	\emph{Max perf.} denotes the performance for a case with the highest performance, $b$ is the dimension of the case with the highest performance (number of nodes is $b^3$), \emph{Max case} denotes the performance for the largest case ($252^3$ nodes).}
\label{tab_perf_spread}
\centering
\begin{tabular}{c c c c}
\hline
Operation              &  Max perf. & $b$ &  Max case \\
\hline
read/write only        & 771        & 220 &   771     \\
propagation only       & 696        & 112 &   650     \\
LBGK incompr.          & 680        & 104 &   632     \\
LBGK q-compr.          & 613        & 100 &   545     \\
LBMRT incompr.         & 468        & 148 &   462     \\
LBMRT q-compr.         & 454        & 152 &   451     \\
\hline
\end{tabular}
\end{table}

\subsubsection{Performance for single precision}
\label{sec_perf_single}

Although the proposed data layout is optimised for double precision representations, we have also analysed its performance for single precision version.
We have measured the performance of six available operations: four LBM kernels and two simplified ones (propagation only and read/write only).
To show an impact of the proposed data layout on the performance we compared the achieved performance to the performance for the standard XYZ layout.
Additionally, for each kernel, we analysed versions with two different numbers of threads per CUDA block: 64 and 128 threads.
Finally, for each combination of the parameters mentioned above, we checked two different limits of registers per thread: 40 and 48 registers resulting in theoretical occupancy 0.75 and 0.63.
The results for dense geometry cavity3D containing $100^3$ nodes are presented in Table \ref{tab_perf_single}.

\begin{figure}[!t]
\centering{
\subfloat{\includegraphics{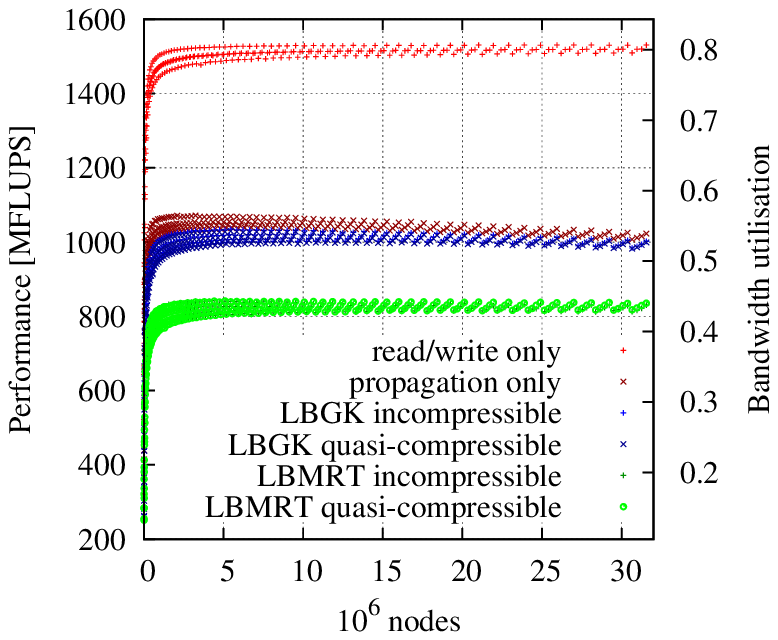}}
\hfill
\subfloat{\includegraphics{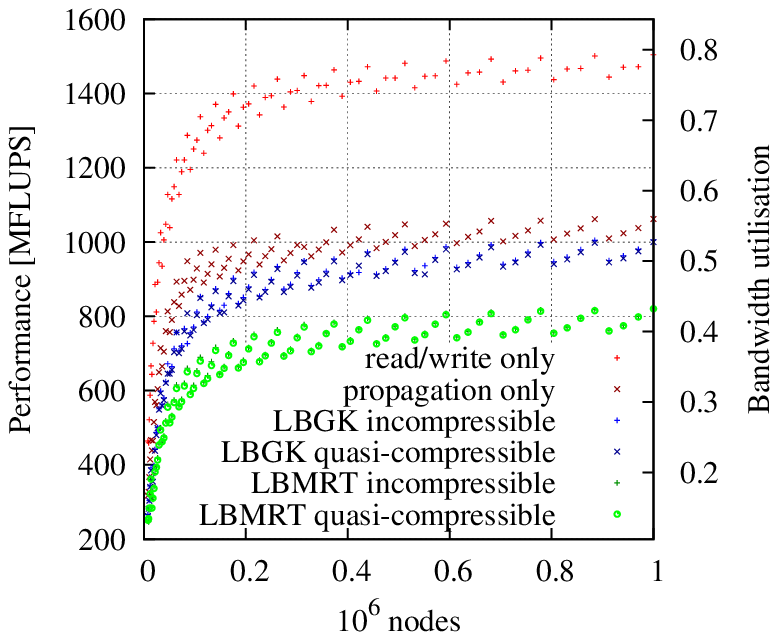}}
}
\caption{
Kernel performance for single precision kernels and cavity3D with geometry size $b^3$ nodes, where $b \in \left< 20, 316 \right>$. 
All measures for XYZ data layout and two tiles per thread block.
}
\label{fig_microflow_mlups_cav3D_single}
\end{figure}

\begin{table*}
\caption{
	Performance in MFLUPS for cavity3D geometry containing $100^3$ nodes and different versions of single precision kernels.
}
\label{tab_perf_single}
\centering
\begin{tabular}{c c c l c c l c c l c c}
\hline
Data layout               & \multicolumn{5}{c}{optimised: XYZ + YXZ + zigzagNE} & & \multicolumn{5}{c}{XYZ}       \\
													\cmidrule{2-6}                      \cmidrule{8-12}
Tiles per block						& \multicolumn{2}{c}{2 tiles}
																				& & \multicolumn{2}{c}{1 tile}
																				   						   & & \multicolumn{2}{c}{2 tiles}
																				   						 								    & & \multicolumn{2}{c}{1 tile} \\
														\cmidrule{2-3}  \cmidrule{5-6}   \cmidrule{8-9}   \cmidrule{11-12}
Registers per thread	    &  48  &  40  & &  48  &  40   & &  48   &  40  & &  48  &  40   \\
\hline                                                                      
read/write only           & 1501 & 1502 & & 1497 & 1498  & & 1503  & \textbf{1504}
																																				 	& & 1497 & 1498  \\
propagation only	        & \textbf{1133}
																 & 1128	& & 1078 & 1050	 & & 1067  & 1066	& & 1037 & 1024  \\
LBGK incompressible	      & 974  & 1003	& & 958	 & 920	 & & 963	 & \textbf{1006}
																																					& & 962	 & 948   \\
LBGK quasi-compressible	  & 968  & 998	& & 958	 & 916	 & & 962	 & \textbf{999}
																																					& & 959	 & 945   \\
LBMRT incompressible	    & 785  & 793	& & 792	 & 777	 & & 810	 & \textbf{823}
																																					& & 805	 & 799   \\
LBMRT quasi-compressible  & 783  & 777	& & 794	 & 779	 & & \textbf{821}
																																	 & 803	& & 806	 & 800   \\
\hline
\end{tabular}
\end{table*}

The presented data show, that the performance of the read/write only kernels is practically the same in all cases and corresponds to bandwidth utilisation $0.79$.
For the propagation only kernel, the proposed data layout gives about 6\% higher performance than for XYZ due to reduced number of memory transactions as shown in Sec. \ref{sec_single_prec_transactions}.
However, when the kernels become more complex, the XYZ layout results in higher performance.
For relatively simple LBGK kernels, the performance for both data layouts is almost the same, but for more complex LBMRT versions the simpler data layout (XYZ) gives up to 5\% better performance.
These results suggest that the computational cost of index calculation for the proposed data layout is not negligible for complex computational models, especially that in the Kepler architecture the same hardware is shared between fixed point and single precision floating point computations \cite{Nvidia:Kep2012}. 
Thus, the results for single precision computations later in this article were obtained for kernels with XYZ data layout and 2 tiles per thread block because these parameters result in the highest performance.
 
The performance of all single precision kernels as a function of a geometry size is shown in Fig. \ref{fig_microflow_mlups_cav3D_single}.
Two main differences may be observed compared to double precision versions shown in Fig. \ref{fig_microflow_mlups_cav3D_double}: much lower bandwidth utilisation for all but the read/write only kernel, and a smaller decrease of performance for increasing geometry size.
Notice also that the performance of both LBGK kernels is similar to the performance of the propagation only kernel.
This shows a small impact of additional computations in single precision LBGK versions, especially additional divisions in quasi-compressible version are much less visible than in double precision (see Fig. \ref{fig_microflow_mlups_cav3D_double}).

\subsection{Propagation performance}

To estimate how the communication between tiles affects the performance we have also measured the performance of propagation for a set of rectangular channels containing about $10^6$ nodes.
The channels differ in dimensions: we have used channels starting from $4 \times 4 \times 62500$ nodes up to $100 \times 100 \times 100$ nodes (1873 combinations in total).
This way allowed us to generate geometries with a different number of faces and edges common to neighbour tiles.

Let $\eta_f$ denote the number of common faces per tile and $\eta_e$ denote the number of common edges per tile.
The values of $\eta_f$ and $\eta_e$ are computed based on the geometric layout, e.g. for four tiles arranged in $2 \times 2 \times 1$ mesh there are 4 common faces and 1 common edge, thus $\eta_f = 4/4$ and $\eta_e = 1/4.$
For channel dimensions used in our measurements the values of $\eta_f$ and $\eta_e$ are the following: channel $4 \times 4 \times 62500$ had about one common face per tile and zero edges, channel $4 \times 8 \times 31248$ had about $1.5$ common face per tile and about $0.5$ common edge per tile, and so forth.
We do not count the numbers of transferred faces/edges because for rectangular channels each face is always common between two tiles and each edge between four, thus the number of data transfers is proportional to $\eta_f$ and $\eta_e$.

\begin{figure}[!b]
\centering
\includegraphics{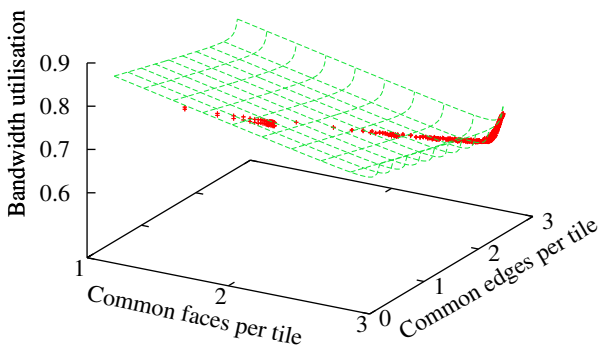}
\caption{
	Performance of propagation as a function of a number of common faces and edges per tile.
	The propagation performance is shown as a fraction of the full GPU memory bandwidth.
	The measurements have been done for rectangular channels containing $10^6$ nodes but with different dimensions.
}
\label{fig_prop_vs_planes_edges_per_tile}
\end{figure}

The propagation performance as a function of common faces per tile and common edges per tile ratios is shown in Fig. \ref{fig_prop_vs_planes_edges_per_tile}.
The highest performance (851 MFLUPS, 90\% GPU memory bandwidth utilisation) was observed for channel $4 \times 62500 \times 4$ nodes.

\begin{table*}[!t]
\caption{
	Propagation performance in MFLUPS for different data layouts inside tile.
	All values for cavity3D with $100^3$ nodes ($25^3$ tiles) and double precision values.
}
\label{tab_perf_propagation}
\centering
\begin{tabular}{c c c c c c c}
\hline
Data layout    & Performance & BW utilisation & L2 reads     &   Memory reads  & Cache hit rate & Instructions \\
\hline
XYZ            &    586      &    $0.618$      & $8~188~609$  &    $7~298~421$  &     $0.109$     &    673       \\
XYZ + zigzagNE &    606      &    $0.639$      & $7~827~219$  &    $6~824~451$  &     $0.128$     &    701       \\
XYZ + YXZ      &    666      &    $0.702$      & $6~730~769$  &    $5~397~210$  &     $0.198$     &    681       \\
all three      &    698      &    $0.736$      & $6~368~191$  &    $4~919~420$  &     $0.228$     &    711       \\
rw only        &    768      &    $0.810$      & $4~812~588$  &    $4~812~524$  &     $\approx 10^{-5}$ &    $~99$ \\
\hline
\end{tabular}
\end{table*}

Most of the points shown in Fig. \ref{fig_prop_vs_planes_edges_per_tile} is placed close to a plane parallel to the "$Bandwidth~utilisation$" axis and defined by a line
\begin{equation}
	\eta_e = 1.85 \cdot \eta_f  - 2.56
	,
	\label{equ_ne_nf}
\end{equation}
which can be treated as a coarse estimation of a ratio of common edges to faces.
Only for small $\eta_f$ ($\eta_f < 2$), which corresponds to very narrow channels of dimension $1 \times k$ tiles, the line defining the plane becomes
\begin{equation}
	\eta_e = \eta_f - 1
	.
\end{equation}

The bandwidth utilisation shown in Fig. \ref{fig_prop_vs_planes_edges_per_tile} can be estimated as 
\begin{equation}
	BU_{p} = 0.92 - \frac{\eta_f}{14.28} - \frac{\eta_e}{25.74} + \frac{0.00104}{ 2.9 - \eta_f} + \frac{0.0023} {2.81 - \eta_e}
	.
\label{equ_bu_prop_est}
\end{equation}
Two main areas can be identified in the dependence shown in Eqn. (\ref{equ_bu_prop_est}).
For $\eta_f < 2.8$ and $\eta_e < 2.6$ the propagation performance falls proportionally to $\eta_f$ and $\eta_e$.
Also, the $\eta_f$ is almost twice as important as $\eta_e$.
For $\eta_f > 2.8$ and $\eta_e > 2.6$ the propagation performance increases in inverse proportion to $\eta_f$ and $\eta_e$.
In this range $\eta_e$ is more important than $\eta_f$, what means that the more square channels (with width similar to depth) allow to achieve a higher performance.
The highest bandwidth utilisation in this range was observed for channel $100 \times 100 \times 100$ nodes (73\%, 696 MFLUPS).

\subsection{Data layout impact}

To measure the performance gains caused by tile memory layouts presented in Sec. \ref{sec_tiling_memory_layout} we have used \emph{nvprof} utility to profile the four propagation kernels with memory layouts XYZ, XYZ + zigzagNE, XYZ + YXZ, and all three (XYZ + YXZ + zigzagNE).
The assignment of memory layouts to $f_i$ data blocks is defined in Sec. \ref{sec_tiling_memory_layout}.
For cases XYZ + zigzagNE and XYZ + YXZ the default memory layout is XYZ.
The results are shown in Table \ref{tab_perf_propagation}.

As a test case, we have used the geometry containing $100 \times 100 \times 100$ nodes due to its regular structure that allows to exactly compute the number of transactions.
The minimal number of write transactions is $25^3~\lbrack tiles \rbrack \cdot 19~\lbrack f_i \rbrack \cdot 16~\lbrack 32-byte~transactions \rbrack = 4~750~000$ 32-byte writes.
The minimal number of 32-byte reads is $4~750~000$ plus transactions for node type reads: $15625 \cdot 64 \cdot 2 / 32 = 62~500$ 32-byte transactions.
The total number of memory transactions is then $4~750~000 + 4~812~500 = 9~562~500$.

The measured numbers of 32-byte memory writes were independent of the used memory layout.
For all layouts, we have observed $4~750~000$ 32-byte writes,  which is the minimal value.
This is a result of the \emph{gather} pattern, where irregular memory accesses to neighbour nodes/tiles occur only during the read stage.

The data from Table \ref{tab_perf_propagation} show that the proposed memory layout ("all three" row) allows achieving the number of memory transactions (both read and write) only  $1.12\%$ larger than the minimal.
This is a result of both a reduction of uncoalesced memory reads (L2 reads column) and an increase of cache hit rate.
Notice also that for "rw only" kernel the numbers of cache and device memory transactions is minimal, what shows the correctness of our theoretical estimations of numbers of memory transactions.

Since the YXZ layout is applied to the largest number of $f_i$ data blocks (eight), its use results in the largest performance increase ($13.65$\% compared with 586 MFLUPS for XYZ layout).
The zigzagNE layout is used only for two $f_i$ data blocks and gives exactly four times smaller performance increase ($3.41$\%).
Notice that the performance increase does not depend on address computation complexity.
Though zigzagNE layout requires $14$ times more additional instructions per $f_i$ data block than YXZ layout (14 compared with 1 instruction per $f_i$ data block), the performance increase scales linearly only with the number of used data blocks regardless of the address computation complexity.

The reductions of the numbers of transactions (for both cache and device memories) for the zigzagNE and YXZ layouts add up: $(8188609 - 7827219) + (8188609 - 6730769) \approx (8188609 - 6368191)$ and $(7298421 - 6824451) + (7298421 - 5397210) \approx (7298421 - 4919420)$.
This means that despite the parallel execution of kernels for $14~\lbrack multiprocessors \rbrack \cdot 16~\lbrack tiles \rbrack$ and shared L2 cache and device memories there is no visible dependence between accesses to different $f_i$ data blocks.

\subsection{Performance for sparse geometries}

\begin{table*}
\centering
\caption{
	Kernels performance in MFLUPS for arrays of random arranged spheres (similar to \cite{Huang:Mul2014}) with porosities between 0.1 and 0.9.
	Geometry size is $192^3$ nodes.
	Results from \cite{Huang:Mul2014} are for Tesla K20 GPU and kernel similar to our LBGK incompressible.
}
\label{tab_perf_ras}
\begin{tabular}{c c c c c c c c c c}
\hline
Geometry porosity      &  0.9  &  0.8  &  0.7  &  0.6  &  0.5  &  0.4  &  0.3  &  0.2  &  0.1  \\
\hline
Tile utilisation       & 0.970 & 0.936 & 0.899 & 0.857 & 0.813 & 0.762 & 0.699 & 0.626 & 0.512 \\
\hline
Kernel                 & \multicolumn{9}{c}{double precision} \\
\cmidrule{1-1}
read/write only        &  761  &  751  &  740  &  726  &  711  &  693  &  668  &  634  &  569  \\
propagation only       &  672  &  667  &  659  &  650  &  638  &  621  &  597  &  560  &  491  \\
LBGK incompressible     & \textbf{649} 
															 & \textbf{645} 
															 				 & \textbf{638}
																			 				 &  628  &  614  &  594  &  563  &  520  &  446  \\
LBGK quasi-compressible &  558  &  558  &  565  &  564  &  555  &  532  &  499  &  459  &  393  \\
LBMRT incompressible     &  450  &  436  &  422  &  404  &  387  &  366  &  343  &  317  &  275  \\
LBMRT quasi-compressible &  436  &  422  &  407  &  390  &  373  &  354  &  332  &  308  &  270  \\
\cite{Huang:Mul2014}   & \textbf{337} 
															 & \textbf{330} 
															 				 & \textbf{334}
																			 				 &  --   &  --   &  --   &  --   &  --   &   --   \\
											 & \multicolumn{9}{c}{single precision} \\
read/write only        & 1503  & 1474  & 1443  & 1406  & 1364  & 1312  & 1245  & 1151  &  959  \\
propagation only       & 1053  & 1034  & 1013  &  989  &  961  &  927  &  880  &  815  &  706  \\
LBGK incompressible     & \textbf{1008} 
															 & \textbf{982}
															 				 & \textbf{954}
																			 				 &  919	 &  880  &  833  &  775  &  706  &  589  \\
LBGK quasi-compressible & 1005  &  976  &  948  &  912  &  873  &  826  &  767  &  700  &  585  \\
LBMRT incompressible     &  815  &  790  &  766  &  734  &  703  &  667  &  624  &  574  &  496  \\
LBMRT quasi-compressible &  819  &  794  &  766  &  733  &  701  &  663  &  620  &  567  &  484  \\
\cite{Huang:Mul2014}   & \textbf{559} 
															 & \textbf{533} 
															 				 & \textbf{554}
																			 				 &  --   &  --   &  --   &  --   &  --   &   --   \\
\hline
\end{tabular}
\end{table*}

To verify the performance of our implementation for sparse geometries we have run simulations for cases similar to presented in \cite{Huang:Mul2014} and \cite{Nita:GPU2013}.
In Table \ref{tab_perf_ras} there are shown results for nine arrays of random arranged spheres with a diameter of 40 lattice units.
The porosity is defined as a ratio of the number of non-solid nodes to the total number of nodes in the bounding box volume (equal to $192^3$ for cases from Table \ref{tab_perf_ras}).
For the presented cases the number of non-solid nodes varies from 701757 (for porosity $0.1$) up to 6381555 (for porosity $0.9$).
The comparison of bandwidth utilisation is shown in Table \ref{tab_ras_bu}.
Our solution achieves the higher bandwidth utilisation for all geometries with porosities reported in \cite{Huang:Mul2014}.
Notice also that the double precision version of the tile based implementation keeps bandwidth utilisation high even for very low porosities.
For porosity 0.2 the performance 520 MFLUPS corresponds to bandwidth utilisation 0.548, which is higher than results from \cite{Huang:Mul2014} for much higher porosities (0.7 to 0.9).

\begin{table}[!b]
\centering
\caption{
	Comparison of bandwidth utilisation with implementation from \cite{Huang:Mul2014}.
	Results from \cite{Huang:Mul2014} are for Tesla K20 GPU, kernel similar to our LBGK incompressible, and random arranged spheres geometry.
}
\label{tab_ras_bu}
\begin{tabular}{l c c c c}
\hline
\multicolumn{2}{l}{Porosity}                              &  0.9  &  0.8  &  0.7  \\
\hline
\multirow{2}{*}{double precision} & this                  & 0.684 & 0.680 & 0.673 \\
                                  & \cite{Huang:Mul2014}  & 0.493 & 0.482 & 0.488 \\
\multirow{2}{*}{single precision} & this                  & 0.531 & 0.518 & 0.503 \\
                                  & \cite{Huang:Mul2014}  & 0.409 & 0.390 & 0.405 \\
\hline
\end{tabular}
\end{table}

\begin{figure}[!b]
\centering
\includegraphics{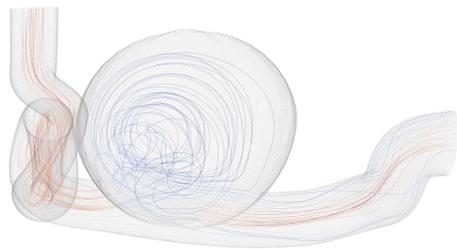}
\caption{Flow pattern within the cerebral aneurysm model.}
\label{fig_aneurysm}
\end{figure}

The second test was the blood flow in a cerebral an\-eu\-rysm (see Fig. \ref{fig_aneurysm}).
We used the model similar to presented in \cite{Huang:Mul2014}, but since our implementation does not support multi-GPU configurations, we reduced the resolution to $730 \times 342 \times 334$ nodes ($84.6 \times 10^6$ nodes in total - the geometry was extended due to tile size, $14.8 \times 10^6$ non-solid nodes, geometry porosity $0.175$, average tile utilisation $\eta_t = 0.931$).
The results are shown in Table \ref{tab_perf_aneuysm}.

For the cerebral aneurysm case, our implementation achieves significantly higher GPU memory bandwidth utilisation than \cite{Huang:Mul2014}.
This is due to the high tile utilisation resulting from a good spatial locality of geometry.
Notice also that the numbers from Table \ref{tab_perf_aneuysm} are close to the values achieved for dense geometries (the right column in Table \ref{tab_perf_spread}).
It shows that when the tile utilisation is high, then the performance of our implementation is almost not dependent on geometry sparsity.

\begin{table}[t]
\centering
\caption{
	Kernels performance for cerebral aneurysm model similar to \cite{Huang:Mul2014}.
	Results from \cite{Huang:Mul2014} are for single precision floating point numbers, four Tesla C1060 GPUs and kernel similar to our LBGK incompressible.
}
\label{tab_perf_aneuysm}
\begin{tabular}{c c c }
\hline
Kernel                &  MFLUPS &  BW utilisation  \\
\hline
\multicolumn{3}{c}{double precision} \\
read/write only       &  753   &  $0.794$   \\
propagation only      &  654   &  $0.689$   \\
LBGK incompr.         &  \textbf{637}   &  \textbf{0.671}   \\
LBGK q-compr.         &  553   &  $0.583$   \\
LBMRT incompr.        &  439   &  $0.463$   \\
LBMRT q-compr.        &  422   &  $0.445$   \\
\multicolumn{3}{c}{single precision} \\
read/write only       &  1480   &  $0.780$   \\
propagation only      &  1021   &  $0.538$   \\
LBGK incompr.         &  \textbf{971}   &  \textbf{0.512}   \\
LBGK q-compr.         &  968   &  $0.510$   \\
LBMRT incompr.        &  789   &  $0.416$   \\
LBMRT q-compr.        &  790   &  $0.416$   \\
\cite{Huang:Mul2014}  &  \textbf{1090}   &  \textbf{0.404}   \\
\hline
\end{tabular}
\end{table}

\begin{figure}[!b]
\centering
\includegraphics{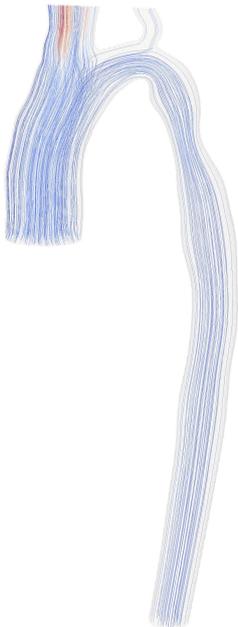}
\caption{Flow pattern within the model of aorta with coarctation.}
\label{fig_coarctation}
\end{figure}

Finally, we have measured the performance for aorta with coarctation case (see Fig. \ref{fig_coarctation}) similar to presented in \cite{Nita:GPU2013}.
The performance comparison is shown in Table \ref{tab_perf_coarctation}.
Since this geometry is rather small (resolution $93 \times 161 \times 442$, about $0.66 \times 10^6$ non-solid nodes, porosity $0.094$, average tile utilisation $\eta_t = 0.807$), the performance of our implementation is up to 14\% lower than for the cerebral aneurysm case shown in Table \ref{tab_perf_aneuysm}.
Small geometry size degrades performance not only due to the low number of nodes (as shown in Fig. \ref{fig_microflow_mlups_cav3D_double}) but primarily due to the reduced tile utilisation (equal to $0.807$ in this case).
Despite this performance degradation, the tile based solution achieves significantly higher memory bandwidth utilisation than implementation presented in \cite{Nita:GPU2013}.

\begin{table}[btp]
\centering
\caption{
	Kernels performance for aorta model with coarctation (similar to \cite{Nita:GPU2013}).
	Results from \cite{Nita:GPU2013} are for GTX 680 GPU, D3Q15 lattice and kernel similar to our LBGK quasi-compressible.
}
\label{tab_perf_coarctation}
\begin{tabular}{c c c }
\hline
Kernel                &  MFLUPS    &  BW utilisation  \\
\hline
\multicolumn{3}{c}{double precision} \\
read/write only       &  $~~~714$   &  $0.753$   \\
propagation only      &  $~~~647$   &  $0.682$   \\
LBGK incompr.         &  $~~~617$   &  $0.650$   \\
LBGK q-compr.         &  $~~~\mathbf{551}$   &  $\mathbf{0.581}$   \\
LBMRT incompr.        &  $~~~382$   &  $0.403$   \\
LBMRT q-compr.        &  $~~~368$   &  $0.388$   \\
\cite{Nita:GPU2013}   &  $\mathbf{\sim 150}$   & $\mathbf{\sim 0.2}$   \\
\multicolumn{3}{c}{single precision} \\
read/write only       &  $~~1393$   &  $0.734$   \\
propagation only      &  $~~~990$   &  $0.522$   \\
LBGK incompr.         &  $~~~864$   &  $0.455$   \\
LBGK q-compr.         &  $~~~\mathbf{857}$   &  $\mathbf{0.452}$   \\
LBMRT incompr.        &  $~~~685$   &  $0.361$   \\
LBMRT q-compr.        &  $~~~681$   &  $0.359$   \\
\hline
\end{tabular}
\end{table}

\subsubsection{Tile utilisation impact}

\begin{figure}[tb]
\centering{
\subfloat{\includegraphics{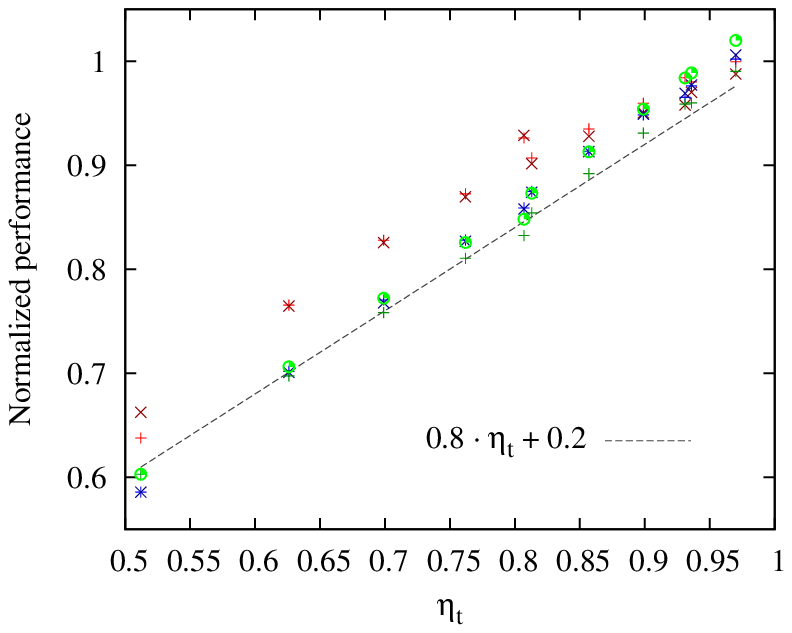}}
\hfill
\subfloat{\includegraphics{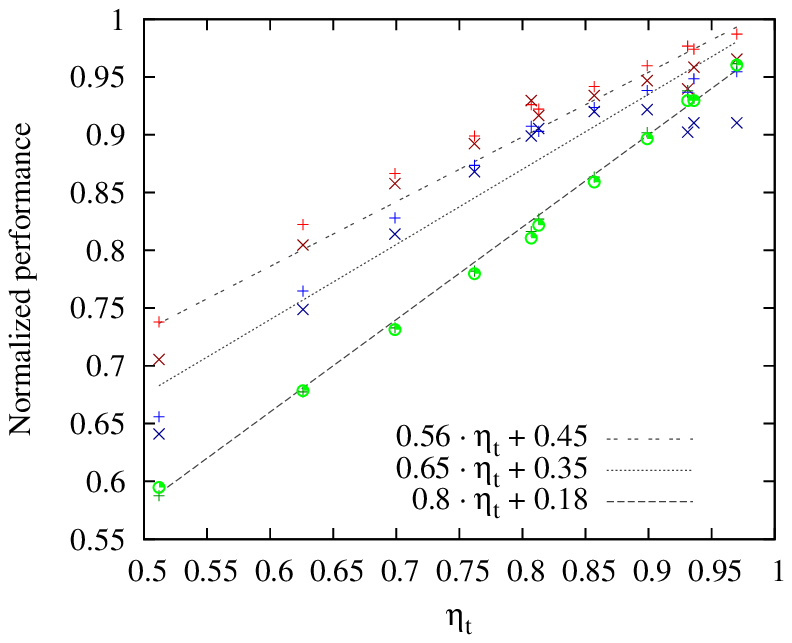}}
}
\caption{
	Normalized performance for all single (top) and double (bottom) precision kernels as a function of the average tile utilisation for all sparse geometries.
	Markers are the same as in Fig. \ref{fig_microflow_mlups_cav3D_double} and \ref{fig_microflow_mlups_cav3D_single}.
}
\label{fig_normalized_perf}
\end{figure}

To show that the performance for sparse geometries depends only on tile utilisation, not on the geometry porosity, we have analysed how the performance scales with $\eta_t$.
The results are shown in Fig. \ref{fig_normalized_perf}.
Performance of all kernels was scaled by the maximum performance observed for the dense cavity3D geometry.
The performance of double precision kernels was divided by the maximum values from Table \ref{tab_perf_spread}, the single precision performance was scaled by bold values from "XYZ 2 tiles" column shown in Table \ref{tab_perf_single}.
For example, if double (single) precision performance of LBGK incompressible kernel for cerebral aneurysm geometry is 637 (971) MFLUPS (Table \ref{tab_perf_aneuysm}) then the scaled performance shown in Fig. \ref{fig_normalized_perf} equals to $637/680 = 0.937$ and $971/1006 = 0.965$.

The graphs presented in Fig. \ref{fig_normalized_perf} show that the performance for sparse geometries depends almost linearly on average tile utilisation (for kernels with the smallest amount of computations the dependence slightly differs from the linear one).
Only a few LBGK simulations for double precision and high tile utilisation drop below lines $\alpha \cdot \eta_t$.
The normalized performance drops proportionally to $\eta_t$ with a ratio $\alpha \cdot \eta_t, \alpha < 1$.
For kernels with high computational cost (LBMRT kernels for double precision and all LBM kernels for single precision, as shown in Sec. \ref{sec_perf_single}) the proportionality constant equals to $0.8$.
When kernels contain a small number of computations, the performance decreases more slowly ($\alpha \approx 0.6$).
The slowest performance drop with $\eta_t$ was observed for the read/write only kernel.

\begin{figure}[tb]
\centering
\includegraphics{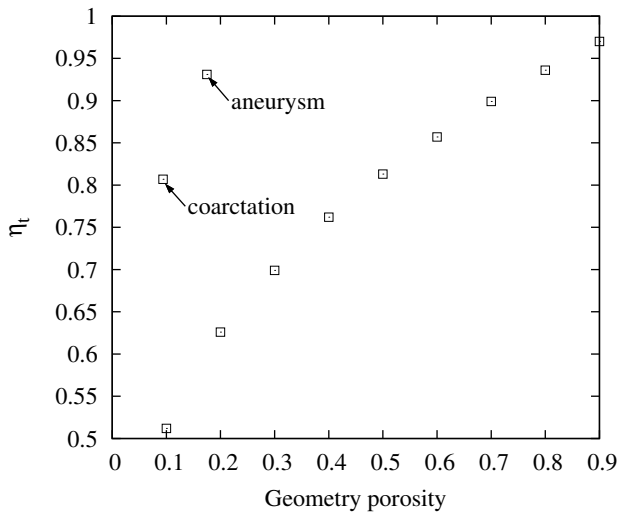}
\caption{
	Average tile utilisation as a function of the geometry porosity for all sparse geometries.
}
\label{fig_tile_utils}
\end{figure}

We have also analysed how the average tile utilisation factor behaves for analysed sparse geometries.
Fig. \ref{fig_tile_utils} shows $\eta_t$ and geometry porosity for random arranged spheres, cerebral aneurysm and aorta with coarctation geometries.
Though for random arranged spheres the value of $\eta_t$ is strongly decreasing with geometry porosity, it can not be treated as a general rule.
Notice that for both blood flow geometries with very low porosity the values of $\eta_t$ stay high.
In practice, we are not able to give a simplified method, thus the tile utilisation must be calculated for each geometry separately.
However, since the tiling of any geometry can be done very efficiently (see Sec. \ref{sec_tiling_overview}), this is not a serious disadvantage.

\section{Conclusions}
\label{sec_conclusions}

In this paper, the GPU based LBM implementation for fluid flows in sparse geometries is presented.
In contrary to the previous solutions for sparse geometries, which use the indirect node addressing, our implementation covers the geometry with the uniform mesh of cubic tiles. 
This method allows to almost completely remove additional data (only a very small amount of data per tile must be used).
Due to the fixed tile size, we have been able to carefully arrange data inside a tile what allowed to additionally decrease the GPU memory traffic.
An important advantage of the proposed solution is that the performance depends only on the average tile utilisation, not on the real geometry porosity.
Since the average tile utilisation can be maintained at high levels even for very sparse geometries, especially these with good spatial locality (for two presented blood flow geometries with porosity 0.175 and 0.094 the average tile utilisation was 0.931 and 0.807), then the proposed data layout allows to simulate flow through sparse geometries with performance close to the recent implementations for dense geometries. 

Our implementation allowed us to achieve up to 680 MFLUPS ($71.7$\% utilisation of maximum theoretical memory bandwidth) on GTX Titan for D3Q19 lattice, LBGK incompressible model and double precision data.
We have also examined the performance for both the LBGK and LBMRT collision models in incompressible and quasi-com\-press\-ible versions.
The performance comparison with existing implementations shows that in all real cases our solution allows to achieve significantly higher performance.
Especially promising results were obtained for the large geometry with good spatial locality (cerebral aneurysm model with $14.8 \times 10^6$ non-solid nodes and porosity 0.175), where for our double precision kernel we observed almost 94\% of peak performance achieved for dense geometry.
This corresponds to $1.66 \times$ higher memory bandwidth utilisation than the highly optimised, single precision implementation based on indirect node addressing.
Even for the very sparse geometries with weak spatial locality (array of random arranged spheres with porosity $0.2$), our double precision implementation has $1.11 \times$ higher bandwidth utilisation than indirect node addressing version for almost dense geometry with porosity 0.9.

For single precision computations, the presented tile based solution offers lower performance gains than for double precision due to lower bandwidth utilisation caused by uncoalesced memory transactions.
Also, the presented data layout optimised for double precision values offers no advantages for single precision compared to the standard, row ordered storage of data inside tiles.
Nonetheless, for all compared geometries the performance of tile based implementation for single precision was also higher than for indirect addressing based ones.

Future work includes the extension to a multi-GPU version to increase the size of supported geometries and the search for a method to increase the tile utilisation (e.g. by a non-uniform tile placement and/or a decrease of the tile size).
We are also going to find memory layouts better suited for single precision values.

\section*{Acknowledgments}

The authors are very grateful to the reviewer’s valuable comments that improved the manuscript.
This work was supported by Wrocław University of Science and Technology, Faculty of Electronics, Chair of Computer Engineering statutory funds.

\end{document}